\documentclass[
 reprint,
 amsmath,amssymb,
 aps,
]{revtex4-2}

\usepackage{graphicx}
\usepackage{dcolumn}
\usepackage{bm}
\usepackage{caption}
\usepackage{subfigure}
\usepackage{float}
\usepackage[braket,qm]{qcircuit}
\usepackage{algorithmic}
\usepackage{algorithm}
\usepackage{colonequals}

\begin{document}

\preprint{APS/123-QED}

\title{Classical Post-processing for Unitary Block Optimization Scheme to Reduce the Effect of Noise on Optimization of Variational Quantum Eigensolvers}

\author{Xiaochuan Ding}
\email{xd13@illinois.edu}
\author{Bryan K. Clark}
\email{bkclark@illinois.edu}
\affiliation{Physics Department, University of Illinois at Urbana-Champaign}

\date{\today}

\begin{abstract}
Variational Quantum Eigensolvers (VQE) are a promising approach for finding the classically intractable ground state of a Hamiltonian. The Unitary Block Optimization Scheme (UBOS) is a state-of-the-art VQE method which works by sweeping over gates and finding optimal parameters for each gate in the environment of other gates.  UBOS improves the convergence time to the ground state by an order of magnitude over Stochastic Gradient Descent (SGD). It nonetheless suffers in both rate of convergence and final converged energies in the face of highly noisy expectation values coming from shot noise. Here we develop two classical post-processing techniques which improve UBOS especially when measurements have large shot noise. Using Gaussian Process Regression (GPR), we generate artificial augmented data using original data from the quantum computer to reduce the overall error when solving for the improved parameters. Using Double Robust Optimization plus Rejection (DROPR), we prevent outlying data which are atypically noisy from resulting in a particularly erroneous single optimization step thereby increasing robustness against noisy measurements. Combining these techniques further reduces the final relative error that UBOS reaches by a factor of three without adding additional quantum measurement or sampling overhead. This work further demonstrates that developing techniques which use classical resources to post-process quantum measurement results can significantly improve VQE algorithms. 

\end{abstract}

\maketitle

\section{Introduction}

In the near term, quantum computers are limited by qubit coherence and gate fidelity. These early noisy intermediate-scale quantum (NISQ) devices \cite{NISQ} have too few physical qubits with low coherence time to implement robust error correction schemes, making them unsuitable for many promising quantum algorithms such as Shor's algorithm \cite{Shor_1,Shor_2,Shor_3,Shor_4,Shor_5,Shor_6,Shor_7,Shor_8} and Grover's algorithm \cite{Grover_1,Grover_2,Grover_3,Grover_4,Grover_5,Grover_6}. To avoid these issues, hybrid classical-quantum algorithms like the quantum approximate optimization algorithm (QAOA) \cite{QAOA} and the variational quantum eigensolver (VQE) \cite{VQE_proposal_1,VQE_proposal_2, VQE_review} leverage the resources of a quantum computer to simulate and sample from a classically intractable state while using classical resources to reduce the demand on qubits and coherence. 

VQE aims to compute an upper bound for the ground-state energy of a Hamiltonian $\hat{H}$, which is generally the first step in computing the properties of molecules and materials \cite{VQE_application_1, VQE_application_2, VQE_application_3, VQE_application_4}. Starting with an \textit{ansatz} which is a quantum circuit built with a set of parametrized quantum gates to model a trial wavefunction, $\ket{\Psi}$, VQE iteratively optimizes the gate parameters of the ansatz to minimize the energy of the trial state by computing the expectation values of operators in the Hamiltonian through measurements on the quantum computer; and then classically updating the gate parameters. 

Given the stochastic nature of measurement on quantum devices \cite{NielsenAndChuang}, one must measure enough copies (denoted as shots) of the same circuit to achieve a given level of precision since the distribution of measurement outcomes on a Hermitian operator has error inversely proportional to square root of the number of shots per circuit \cite{shot_scaling}. It is worth noting that this error scaling can be improved by using more sophisticated quantum algorithms such as quantum phase estimation \cite{QPE}; unfortunately, these techniques require much deeper circuits making them impractical for their current generation of quantum devices. 

The standard approach to VQE has been improved in various ways including ansatz construction \cite{VQE_previous_HEA_1,VQE_previous_HEA_2,VQE_proposal_1,VQE_previous_UCC_1,VQE_previous_UCC_2,VQE_previous_SP_1,VQE_previous_SP_2,VQE_previous_ADAPT_1}, efficient measurement strategy \cite{VQE_previous_grouping_1,VQE_previous_grouping_2,VQE_previous_grouping_3,VQE_previous_grouping_4,VQE_previous_QOT_1,VQE_previous_QST_1,VQE_previous_QST_2,VQE_previous_CS_1,VQE_previous_CS_2,VQE_previous_CS_3,VQE_previous_CS_4,VQE_previous_CS_5}, error mitigation techniques \cite{QEM_ZNE_1,QEM_ZNE_2,QEM_ZNE_3,QEM_ZNE_4,QEM_ZNE_5,QEM_ZNE_6,QEM_ZNE_7,QEM_ZNE_8,QEM_PEC_1,QEM_PEC_2,QEM_PEC_3,QEM_MNM_1,QEM_LB_1,QEM_LB_2}, and optimization strategies \cite{VQE_previous_GD_BFGS_1,VQE_previous_GD_BFGS_2,VQE_previous_GD_BFGS_3,VQE_previous_GD_BFGS_4,VQE_previous_GD_NGD_1,VQE_previous_GD_NGD_2,VQE_previous_GD_NGD_3,VQE_previous_GF_NM_1,VQE_previous_GF_Powell_1,VQE_previous_GF_analytical_1,VQE_previous_GF_analytical_2}. Classical machine learning techniques such as Koopman Operator Learning \cite{DiLuo} and physics-informed neural network \cite{PINN} have also been used to improve VQE. Typical optimization strategies for VQE algorithms are gradient-free classical optimization methods including Nelder-Mead method and Powell's algorithm \cite{VQE_previous_GF_NM_1,VQE_previous_GF_NM_2, VQE_previous_GF_Powell_1}, gradient-based searching strategy \cite{VQE_previous_GD_NGD_3,VQE_previous_HEA_1,VQE_previous_ADAPT_1,GD_example_1,GD_example_2,GD_example_3,GD_example_4,GD_example_5,GD_example_6,GD_example_7,GD_example_8,GD_example_9,GD_example_10,GD_example_11,GD_example_12}, and analytical methods such as Anderson Acceleration \cite{VQE_previous_GF_analytical_1,VQE_previous_GF_analytical_1}. 
Traditional methods such as Stochastic Gradient Descent (SGD) face several challenges including local minima, significant hyperparameter tuning, slow convergence, and exponentially vanishing gradients. 

The unitary block optimization scheme (UBOS) is a gradient-free and hyperparameter-free optimization algorithm \cite{UBOS}. By optimizing a subset of parameters at each step using the effective Hamiltonian $\tilde{H}$, it avoids gradient calculation, tunnels through some local minima, makes nontrivial steps decreasing the energy when facing barren plateaus, and requires an order of magnitude less expectation value measurements than Stochastic Gradient Descent (SGD) \cite{UBOS}. 

The key step of UBOS is to generate the effective Hamiltonian for a gate in a fixed environment. The standard approach to accomplish this is to directly measure the matrix elements using a separate quantum circuit for each element;  this approach we refer to as \textit{\textbf{D-UBOS}}. An alternative approach is to infer the effective Hamiltonian from pairs of gate parameters and their corresponding energies $\{(t_j,E_{\textrm{measured}})\}$, which we will refer to as \textit{\textbf{E-UBOS}}. 

One of the primary concerns for VQE algorithms is its shot budget (total amount of measurements). One approach to reducing the shot budget is to simply take fewer measurements per circuit resulting in much larger stochastic errors due to the finite number of shots (denoted as shot noise). Significant shot noise often hinders classical optimizers from finding the true global minimum; many methods including UBOS plateau at an energy level above the optimal VQE energy of the ansatz (VQEOPT). This raises scalability concerns for the VQE algorithm \cite{VQE_noise_1,VQE_noise_2,VQE_noise_3}. 

There are efforts on reducing the number of shots required for gradient-based VQE such as estimating the gradient with few-shot measurements by parameter shift rule \cite{SGD}, modifying the number of shots for estimating each component of the gradient using an adaptive optimizer \cite{iCANS} and by bootstrapping and resampling based on the variance of obtained shots \cite{VAQC}, and importance sampling \cite{importance_sampling_1,importance_sampling_2}. These methods aim to frugally select the number of shots while still using SGD.

For UBOS, shot noise in quantum measurements causes error in the effective Hamiltonian $\tilde{H}$ which leads to inaccurate state energy estimation using $\tilde{H}$. D-UBOS has no measure against error in measurement outcomes. In E-UBOS one can partially mitigate shot noise by increasing the number of $(t_j,E_{\textrm{measured}})$ pairs at the cost of additional quantum measurements. Unfortunately, when we empirically compare, E-UBOS as naively formulated still requires the same amount or slightly more shot budget to reach the same energy error as D-UBOS (see Fig.~\ref{figs:testing all methods}). 

In this paper, we develop generalizations of E-UBOS to resolve this problem. Our philosophy is that while the classical optimization of the gate parameters for the exact effective Hamiltonian is straightforward, classical shot-noise-aware post-processing techniques can help reach a better energy without taxing the shot budget, especially when the quantum measurements are very noisy. We introduce two techniques: Data Augmentation with Gaussian Process Regression (GPR) and Double Robust Optimization plus Rejection (DROPR). We demonstrate that these two classical post-processing techniques can effectively suppress shot noise in quantum measurements and reduce the relative energy error of the full optimization roughly by a factor of 3 for all choices of hyperparameters in the range studied.

The rest of the paper is organized as follows: In Sec.~\ref{section:intro to UBOS methods} we briefly review how to implement different types of UBOS. Next, in Sec.~\ref{section:intro to classical post-processing}, we describe the classical post-processing techniques. Then in Sec.~\ref{section:tests} we benchmark the performance of E-UBOS with these techniques applied and compare it with D-UBOS. Finally, we conclude the paper in Sec.~\ref{section:discussion} with a discussion of our main results.

\section{Introduction to UBOS methods}
\label{section:intro to UBOS methods}
\subsection{Review of UBOS}

In this paper we describe all types of UBOS using a variational ansatz, 
\begin{equation}
\label{eq:UBOS ansatz}    
\ket{\psi}=\prod_{j=1}^{K}{U_j\ket{0}}
\end{equation}
obtained by applying K generic two-qubit unitaries $U_j\in{SU(4)}$ (i.e., quantum gates), to adjacent qubits in a brickwork pattern with gate depth \textit{d}. The generic two-qubit unitary, \textit{$U_j$}, can be written as a linear combination of 16 two-qubit Pauli strings,
\begin{equation}
\label{eq:UBOS generic unitary}    
U_j=\sum_{\alpha,\beta=0}^{3}{t_j^{\alpha\beta}P^{\alpha\beta}}
\end{equation}
where $P^{\alpha\beta}=\sigma^\alpha \otimes \sigma^\beta$, $\sigma^{\alpha,\beta}\in\{I, X, Y, Z\}$ are Pauli matrices, and the complex coefficient $t_j^{\alpha\beta}$ are constrained to preserve the unitarity of $U_j$ (see Appendix~\ref{appendix:KAK}).

UBOS then parameterizes the state by the gate parameters $\{ \mathrm{ t_{1}, t_{2}...t_{K} } \}$ where $\mathrm{ t_{j}\equiv(t_j^{00},t_j^{01}...t_j^{33})}$,
\begin{align}
\label{eq:UBOS parameterized state}    
\ket{\psi} & = (\prod_{k=1}^{j-1}{U_k})U_j(\prod_{k=j+1}^{K}{U_k})\ket{0} \nonumber \\
           & = (\prod_{k=1}^{j-1}{U_k})(\sum_{\alpha,\beta=0}^{3}{t_j^{\alpha\beta}P^{\alpha\beta}})(\prod_{k=j+1}^{K}{U_k})\ket{0} \nonumber \\
           & = \sum_{\alpha,\beta=0}^{3}{t_j^{\alpha\beta}\ket{\psi_j^{\alpha\beta}}}
\end{align}
where 
\begin{equation}
\label{eq:UBOS state derivative}    
\ket{\psi_j^{\alpha\beta}}=(\prod_{k=1}^{j-1}{U_k})P^{\alpha\beta}(\prod_{k=j+1}^{K}{U_k})\ket{0}
\end{equation}
is the result of substituting gate $U_j$ by the Pauli operator $P^{\alpha\beta}$ and has the same circuit depth $d$. It is also the partial derivative of $\ket{\psi}$ with respect to $t_j^{\alpha\beta}$ which we use to estimate the gradient in SGD.

Given some Hamiltonian $\mathrm{\hat{H}}$, keeping the parameters for all but the \textit{j}th gate fixed, UBOS writes the energy as a function of the \textit{j}th gate parameters,
\begin{align}
E(\textbf{t}_j) & = \bra{\psi}\hat{H}\ket{\psi} \nonumber \\
                & = \left(\sum_{\alpha',\beta'=0}^{3}{t_j^{*\alpha'\beta'}\bra{\psi_j^{\alpha'\beta'}}}\right)\hat{H}\left(\sum_{\alpha,\beta=0}^{3}{t_j^{\alpha\beta}\ket{\psi_j^{\alpha\beta}}}\right) \nonumber \\
                & = \sum_{\alpha,\beta,\alpha',\beta'=0}^{3}{t_j^{*\alpha'\beta'} \tilde{H}^{\alpha'\beta';\alpha\beta} t_j^{\alpha\beta}} 
                \label{eq:UBOS energy sum form}
                \\
                & = \textbf{t}_j^\dag \tilde{H} \textbf{t}_j \label{eq:UBOS energy}
\end{align}
with the effective Hamiltonian for gate $j$,  $\tilde{H}$ as
\begin{equation}
\label{eq:effective matrix elements}    
\tilde{H}^{\alpha'\beta';\alpha\beta} = \bra{\psi_j^{\alpha'\beta'}}\hat{H}\ket{\psi_j^{\alpha\beta}}.
\end{equation}
$\tilde{H}$ is a $16\times16$ Hermitian matrix with 256 unique real parameters for its matrix elements (136 for real component and 120 for imaginary component), and is independent of $t_j$.

Once $\tilde{H}$ is obtained, UBOS classically optimizes the gate parameters for the \textit{j}th gate while keeping all other gates fixed by minimizing Eq.~\ref{eq:UBOS energy} with respect to the gate parameters $\textbf{t}_j$ under the unitary constraint (see Appendix~\ref{appendix:KAK}). This is a 16-parameter optimization problem that can be solved using any classical technique such as gradient descent, Nelder-Mead, etc. UBOS then sweeps over gates, optimally minimizing the energy of one gate at a time while keeping other gates temporarily ﬁxed. The update order for gates is shuffled to be random for every sweep, a.k.a. epoch.

To obtain the effective Hamiltonian, D-UBOS directly measures these matrix elements with Hadamard test circuits of depth at most $2\textit{d}$ (see Appendix~\ref{appendix:UBOS measurement circuits}), where the Hamiltonian is expanded into the sum of unitary operators. The expectation value measurement of each Hamiltonian component requires (many copies of) a separate circuit and the amount of shot noise depends on the number of shots per circuit (denoted as $n_{\textrm{shots}}$). The total number of measurements $N_{\textrm{meas}}$ scales as $O(dn_q^2 n_{\textrm{shots}})$ (see Appendix~\ref{appendix:total number of measurements}) where $n_q$ is the number of qubits and $d$ is the ansatz depth.

\subsection{Review of E-UBOS}
As described in Appendix G of the original UBOS paper \cite{UBOS}, the matrix elements of \textit{$\tilde{H}$} in Eq.~\ref{eq:UBOS energy} are linear unknowns and independent of the gate parameters at the \textit{$j$}th gate, so instead of measuring them individually, we can solve for them from a system of linear equations obtained by measuring the energies of the states with the \textit{$j$}th gate replaced by a two-qubit unitary generated with different, randomly chosen gate parameters. One advantage of this approach is that the depth of the circuit required for measurement is only d (see Appendix~\ref{appendix:UBOS measurement circuits}). 

By writing $\textit{t}_j$ and \textit{$\tilde{H}$} as their complex form, $t_j=\operatorname{Re}{[t_j]}+i\operatorname{Im}{[t_j]}$ and $\tilde{H}=\operatorname{Re}{[\tilde{H}]}+i\operatorname{Im}{[\tilde{H}]}$, We can rewrite Eq.~\ref{eq:UBOS energy sum form} as 
\begin{multline}   
E(\textbf{t}_j) = \sum_{\alpha',\beta',\alpha,\beta=0}^{3}t_{\textrm{j,R}}^{\alpha'\beta'\alpha\beta}\operatorname{Re}{[\tilde{H}^{\alpha',\beta';\alpha,\beta}]} \\ + t_{\textrm{j,I}}^{\alpha',\beta',\alpha,\beta}\operatorname{Im}{[\tilde{H}^{\alpha',\beta';\alpha,\beta}]} 
\label{eq:alt UBOS energy}
\end{multline} 
where 
\begin{align}
t_{\textrm{j,R}}^{\alpha',\beta',\alpha,\beta} &\equiv \operatorname{Re}{[t_j^{\alpha',\beta'} (t_j^{\alpha,\beta})^*]} \label{eq:t_j^R} \\
t_{\textrm{j,I}}^{\alpha',\beta',\alpha,\beta} &\equiv \operatorname{Im}{[t_j^{\alpha',\beta'} (t_j^{\alpha,\beta})^*]} \label{eq:t_j^I}
\end{align}
are quadratic forms of the $\textit{t}_j$ components (see Appendix~\ref{appendix:E-UBOS math}). Therefore, every random gate parameter vector $\textit{t}_j$ corresponds to a noiseless energy $E$. 

We generate random $t_j$ by sampling a random two-qubit unitary from the unitary Haar measure and then performing decomposition in the Pauli basis. State energies are calculated by the sum of expectation values of $\hat{h}_i$ where $\hat{h}_i$ are the components of the Hamiltonian. Since the Hamiltonian studied contains only local $\{Z, XX, YY, ZZ\}$ operators, by measuring their expectation values on quantum devices (see Appendix~\ref{appendix:UBOS measurement circuits}), we obtain the measured state energy, $E_{\textrm{measured}}$, which is a stochastically noisy \textit{observation} of pair $(t_{j}, E_{\textrm{measured}})$. To characterize shot noise in the measured energy, we can write $E_{\textrm{measured}}$ as
\begin{equation}
E_{\textrm{measured}} = E + \delta E = \textbf{t}_j^\dag \tilde{H} \textbf{t}_j + \delta E \label{eq:UBOS energy with error} 
\end{equation}
where $\delta E$ is the error in the measured state energy due to shot noise (assuming no error from the experimental process of measuring the device).

Given $n_{\textrm{obs}}$ pairs of $(t_j, E_{\textrm{measured}})$, we can form a system of $n_{\textrm{obs}}$ linear equations and determine \textit{$\tilde{H}$} by linear least squares fit over the system. The number of linearly independent components in the set $T_j \equiv \{ t_{\textrm{j,R}}^{\textrm{nm}} \cup t_{\textrm{j,I}}^{\textrm{nm}} \}$ is found to be 226 and is smaller than 256, the number of unique real parameters in the $16\times16$ Hermitian matrix \textit{$\tilde{H}$}, which leads to non-unique effective Hamiltonian that satisfies Eq.~\ref{eq:alt UBOS energy}. However, any effective Hamiltonian that satisfies Eq.~\ref{eq:alt UBOS energy}, or equivalently, Eq.~\ref{eq:UBOS energy}, is suitable for optimization, and by increasing $n_{\textrm{obs}}$ to overconstrain the system, we can add additional robustness against error in measured state energy which is normally reduced by increasing shots per circuit, $n_{\textrm{shots}}$. In other words, to increase the accuracy of the effective Hamiltonian, one can use larger $n_{\textrm{obs}}$ to compensate for the large stochastic sampling noise due to small $n_{\textrm{shots}}$ and vice versa (note that the minimum value of $n_{\textrm{obs}}$ is 226).

\section{Classical post-processing for E-UBOS}
\label{section:intro to classical post-processing}

In the current applications of UBOS, the effective Hamiltonian at each step is computed from a finite number of shots which results in a noisy effective Hamiltonian which we assume to be `exact' when computing the new parameters. Here we suggest an alternative approach which does a significant amount of classical post-processing on the data gained from E-UBOS (and sometimes additional quantum post-processing). This classical post-processing can be aware of the noisy nature of the measurements allowing it to better select new parameters. In this paper, we introduce two techniques: Data Augmentation with Gaussian Process Regression (GPR) and Double Robust Optimization plus Rejection (DROPR) inspired from approaches in machine learning. 

\subsection{Data Augmentation with Gaussian Process Regression}

An E-UBOS optimization step involves three parts: first, it obtains a set of $(t_j, E_{\textrm{measured}})$ pairs through quantum measurements. For convenience, we will refer to this set as the \textit{initial set}, denoted by $S_{\textrm{init}}$. Then, it computes the effective Hamiltonian from a system of linear equations formed with $S_{\textrm{init}}$. Finally, it classically finds the gate parameters that minimizes the state energy based on the obtained effective Hamiltonian. When the number of shots per circuit is small, large shot noise in pairs in $S_{\textrm{init}}$ can cause severe error in the calculated effective Hamiltonian.

One could increase the accuracy of the effective Hamiltonian by having more $(t_j, E_{\textrm{measured}})$ pairs but this would obviously involve a larger shot budget.  An alternative to this approach is to generate artificial pairs using ``Data Augmentation'' \cite{data_augmentation_1, data_augmentation_2, data_augmentation_3} in such a way that expanding $S_{\textrm{init}}$ with this additional artificial data will lead to more accurate estimation of the effective Hamiltonian. Artificial data is created by predicting the energy of new random $t_j$ based on existing observations. Traditional algorithms such as least squares regression suffer from the large shot noise in each observation and the nonlinear relationship between the features (gate parameters) and the target (energy). An alternative approach is to use Gaussian Process Regression, which returns an evidence-based posterior probability distribution over possible functions that fit a set of points \cite{GPR, GPR_example}. Here we describe how GPR can be used to generate new artificial data for our VQE energies including for completeness the underlying theory for how GPR selects the new data. 

We assume that the shot noise in the measured energy follows some Gaussian distribution, $\delta E \sim N(0, \sigma_{\epsilon}^2)$, such that for each random gate parameters $t_j$, the corresponding measured energy follows the normal distribution $E_{\textrm{measured}} \sim N(E, \sigma_{\epsilon}^2)$ where $E$ is the noiseless energy. Then we can model the collection of measured energies as a multi-variate normal (MVN) distribution $P(\{E_{\textrm{measured}}\}|\{t_j\})$. Any sample from this MVN distribution would correspond to a function which is possibly suitable to describe the relationship between the various $t_j$ and $E$. However these functions are very unlikely to be smooth enough for regression purpose and the number of possible functions is infinite. Therefore, we determine the possible functions by sampling with a kernel function which measures the similarity (covariance) between two $t_j$'s, following the logic that similar $t_j$'s should lead to similar $E$'s. This constitutes our prior which is a collection of infinite numbers of smooth functions derived with the kernel and its mean function equals to zero.

We can write a collection of observed data and artificial data as $\{(t_{\textrm{j,obs}},E_{\textrm{measured}})\}$ and $\{(t_{\textrm{j,new}},E_{\textrm{predict}})\}$, respectively, where $E_{\textrm{predict}}$ are unknown. Then we can model $E_{\textrm{measured}}$ and $E_{\textrm{predict}}$ as a MVN distribution in block matrix notation:

\begin{multline}
P(\{E_{\textrm{measured}}\}, \{E_{\textrm{predict}}\} | \{t_{\textrm{j,obs}}\}, \{t_{\textrm{j,new}}\} ) \\
\sim N
\begin{pmatrix}
\begin{bmatrix}
M_{\textrm{obs}}(\{t_{\textrm{j,obs}}\}) \\
M_{\textrm{new}}(\{t_{\textrm{j,new}}\})
\end{bmatrix}
,
\begin{bmatrix}
\hat{K}_{\textrm{obs,obs}} & K_{\textrm{obs,new}}  \\
K_{\textrm{obs,new}}^T & K_{\textrm{new,new}}
\end{bmatrix}
)
\end{pmatrix}
\label{eq:GPR joint distribution}
\end{multline}

where $M_{\textrm{obs}}$ and $M_{\textrm{new}}$ are the mean functions of the MVN distribution for the collection of observed data and artificial data, respectively. $\hat{K}_{\textrm{obs,obs}}=K_{\{t_{\textrm{j,obs}}\},\{t_{\textrm{j,obs}}\}}+\sigma_{\epsilon}^2$ is the covariance matrix between all $t_j$ in the observed data with shot noise added. $K_{\textrm{obs,new}}=K_{\{t_{\textrm{j,obs}}\},\{t_{\textrm{j,new}}\}}$ and $K_{\textrm{new,new}}=K_{\{t_{\textrm{j,new}}\},\{t_{\textrm{j,new}}\}}$ are the covariance matrices between all $t_j$ in the observed data and in the artificial data and between all $t_j$ in the artificial data, respectively.

Now our observations $\{(t_{\textrm{j,obs}}, E_{\textrm{measured}})\}$ become partial observations of this joint normal distribution. Therefore, by Marginal and Conditional Distribution of Multivariate Normal Distribution Theorem, we can find that the conditional probability distribution of the predicted energies follows the MVN in block matrix notation

\begin{multline}
P(\{E_{\textrm{predict}}\} | \{E_{\textrm{measured}}\}, \{t_{\textrm{j,obs}}\}, \{t_{\textrm{j,new}}\} ) \\
\sim N(\mu,\Sigma)
\label{eq:GPR posterior for new data}
\end{multline}
where $\mu = K_{\textrm{obs,new}}^T \hat{K}_{\textrm{obs,obs}}^{-1} \{ E_{\textrm{measured}}\}
$ and $\Sigma = K_{\textrm{new,new}} - K_{\textrm{obs,new}}^T \hat{K}_{\textrm{obs,obs}}^{-1} K_{\textrm{obs,new}} $ are its mean and covariance, respectively. The mean value of each feature of this MVN is then the predicted energy for each artificial $t_j$ with maximum likelihood. 

Equivalently, one can explicitly find the probability distribution of possible functions $f$ instead of energies. In this way, one sees that Gaussian Process uses some kernel function to generate a prior for probability distribution of possible functions and calculates the posterior probability distribution of the functions for the observed data (evidence), $P(\{f_{\textrm{measured}}\} | \{(t_{\textrm{j,obs}}, E_{\textrm{measured}})\})$, which is similar to Bayesian Inference process. Then, it repeats the same formalism as in Eq.~\ref{eq:GPR joint distribution} and Eq.~\ref{eq:GPR posterior for new data} with energy terms replaced by corresponding function terms.

Gaussian Process Regression is more useful in this data augmentation task than least squares regression. It relaxes the form of the predicted model from one function to a probability distribution of possible functions, which is more effective in dealing with the nonlinear relationship between $t_j$ and $E$ and the non-unique effective Hamiltonians that fit the observations well.

We propose the following scheme of performing data augmentation using Gaussian Process Regression (GPR): we assemble overlapping subsets of $(t_j, E_{\textrm{measured}})$ pairs from $S_{\textrm{init}}$, train a Gaussian Process model using Gaussian Process Regressor for each subset, generate artificial $(t_j, E_{\textrm{predict}})$ pairs by applying those models on new random gate parameters $t_j$, and merge them with the $S_{\textrm{init}}$ to create an expanded data set. We choose the radial basis function (RBF) as kernel which is the common default. See Appendix~\ref{appendix:GPR parameters} for a detailed discussion of hyperparameter choices for this technique.

We empirically find that dividing the initial set into overlapping subsets leads to better optimization results than using the whole initial set for GPR model training and artificial data generation despite the subsets involving less total data (see Appendix.~\ref{appendix:GPR parameters}.

In the ideal GPR data augmentation scheme, subsets of pairs will not overlap with each other. However, given the constraint that the minimum size of the subset is 226, if the number of elements in the initial set is not much bigger than 226 due to limited total number of measurements, we have to allow overlap between subsets which leads to non-negligible similarity between the models learned from different subsets of pairs.

\subsection{Double Robust Optimization Plus Rejection}

With a real quantum device, the shot noise in observations and thus in the computed effective Hamiltonian is inevitable, which hinders the classical optimizer from finding the true gate parameters $t_j$ that minimize the state energy. However, we can use robust optimization to mitigate the impact of shot noise. Robust optimization is a widely-applied approach to deal with data uncertainty in optimization that does not require the knowledge on the true probability distribution of uncertain data \cite{robust_optimization_1, robust_optimization_2}. Robust optimization seeks to find solutions that perform well across a range of possible conditions, rather than optimizing for a specific set of conditions.

Similar to the argument in previous section, we assume some of the $(t_j, E_{\textrm{measured}})$ pairs in the initial set $S_{\textrm{init}}$ are extremely noisy and a fit involving these corrupted pairs will give a bad effective Hamiltonian. Given the difficulty of screening them out in advance, we create subsets of $S_{\textrm{init}}$, in each of which a random portion of pairs in $S_{\textrm{init}}$ are dropped out to mitigate their impact. Since some of these subsets are less likely to have the particularly bad pairs, if we fit an effective Hamiltonian $\tilde{H}$ from each subset, some of the $\tilde{H}$'s may be less noisy due to the absence of (at least some of) the bad pairs. Then instead of using the energy calculated using Eq.~\ref{eq:UBOS energy} with one $\tilde{H}$ as loss function for gate parameters optimization, we instead find the gate parameter $t_j$ such that 
\begin{equation}
    \textrm{max}\{ t_j^\dagger \tilde{H}_k t_j : k = 1,2,...,n_{\textrm{subset}}\}
    \label{eq:worst-case robust optimization}
\end{equation}
is minimized where $n_{\textrm{subset}}$ is the number of effective Hamiltonians obtained. This is called a worst-case robust optimization of gate parameters.

Finding the gate parameters that minimizes the energy evaluated with different effective Hamiltonians can reduce the impact of a small portion of extremely noisy pairs in the initial set. However, this strategy fails when there is a particularly bad effective Hamiltonian within the collection of all $\tilde{H}$'s that always gives the worst energy and forces the classical optimizer to accommodate to it. In this case, the gate parameters after optimization may have the state energy calculated with bad $\tilde{H}$ minimized while giving a noiseless state energy worse than before the optimization step. To further reduce the impact of noisy $\tilde{H}$ in the collection of all $\tilde{H}$'s, we add a second layer of robust optimization: after obtaining a collection of $\tilde{H}$'s (denoted as $S_{\tilde{H}}$) from all subsets of pairs, we create sub-collections of $\tilde{H}$'s, in each of which a random portion of $\tilde{H}$'s in $S_{\tilde{H}}$ are dropped out. We use each sub-collection to perform worst-case robust optimizations independently, each of which yields a $t_j$. To select the one that gives the best noiseless energy we do additional quantum measurements for each $t_j$ and the original pre-optimized $t_j$ with more shots per circuit than the shots used for obtaining the observations in $S_{\textrm{init}}$ (see Appendix~\ref{appendix:DROPR energy measurement}). Based on newly measured energies, we choose the best $t_j$ (or reject the optimization move if the original parameters are lowest in energy leaving the gate parameters unchanged). 

\begin{minipage}{0.45\textwidth}
    \small
    \begin{algorithm}[H]
        \small
        \caption{DROPR}
        \label{algorithm:DROPR scheme}
        \begin{algorithmic}
            \STATE\textbf{Input:} Pre-optimized quantum circuit $|\mathrm{\Psi}\rangle$, initial set of $n_{\mathrm{obs}}$ pairs of $(t_j, E_{\mathrm{measured}})$, number of shots per circuit $n_{\mathrm{shots}}$, the index of gate to be optimized $j$, DROPR parameters $n_{\mathrm{subset}}$, $L_{\mathrm{subset}}$, $n_{\mathrm{subcol}}$, $L_{\mathrm{subcol}}$, $n_{\mathrm{dup}}$
            \STATE \textbf{Output:} Optimal parameters $\mathrm{t_j}$
            \STATE Randomly form $n_{\mathrm{subset}}$ overlapping subsets of pairs from the initial set. Each subset has $L_{\mathrm{subset}}$ pairs.
            \FOR{$i = 0$ to $n_{\mathrm{subset}} - 1$}
                \STATE Convert $t_j$ of each pair in this subset to its quadratic form through Eq.~\ref{eq:t_j^R} and Eq.~\ref{eq:t_j^I}
                \STATE Compute $\mathrm{\tilde{H}}$ through linear least square regression using Eq.~\ref{eq:alt UBOS energy}
            \ENDFOR
            \STATE Randomly form $n_{\mathrm{subcol}}$ overlapping sub-collections of $\mathrm{\tilde{H}}$ from the collection of all $\mathrm{\tilde{H}}$ obtained. Each sub-collection has $L_{\mathrm{subcol}}$ elements.
            \FOR{$i = 0$ to $n_{\mathrm{subcol}} - 1$}
                \STATE Find a contender $t_j$ that minimizes Eq.~\ref{eq:worst-case robust optimization}
                \STATE Measure the energy of the state with the $\mathrm{j}$th gate replaced by a two-qubit unitary generated with contender $\mathrm{t_j}$ with $n_{\mathrm{dup}} \times n_{\mathrm{shots}}$ shots per circuit
            \ENDFOR
            \STATE Measure the energy of state $|\mathrm{\Psi}\rangle$ with $n_{\mathrm{dup}} \times n_{\mathrm{shots}}$ shots per circuit
            \STATE Select the optimal $t_j$ with the best measured energy (or reject the change if the original state energy is optimal)
        \end{algorithmic}
    \end{algorithm}
\end{minipage}

The full description of the scheme is shown in Algorithm.~\ref{algorithm:DROPR scheme}. This strategy bears resemblance to Median-of-Means trick used in Classical Shadows \cite{median_of_means_1, median_of_means_2} and Metropolis-Hastings Algorithm in Monte Carlo methods and is effective in preventing the algorithm from accepting gate parameters which give plausible state energy calculated with noisy effective Hamiltonian but has noiseless state energy worse than before the optimization step.

Again, in the ideal scheme there shouldn't be overlap between subsets of pairs for effective Hamiltonian fitting. However, given the constraint of minimum size of the subset of pairs being 226, if the size of initial set is small, we have to allow overlap between subsets.

\subsection{GPR and DROPR combined}

For the rest of the paper, we will refer to the E-UBOS method with GPR technique as \textbf{\textit{Eg-UBOS}} and the E-UBOS method with DROPR technique as \textbf{\textit{Ed-UBOS}}. Since the GPR technique focuses on expanding the measured dataset to be more comprehensive for the effective Hamiltonian computation and the DROPR technique aims to improve the search for optimal gate parameters given some set of noisy data, we can apply these two techniques in a combined way to take their complementary advantages.The E-UBOS method with both techniques applied is called \textbf{\textit{Edg-UBOS}}.

\section{Numerical Comparisons of Approaches}
\label{section:tests}
\subsection{Comparing Edg-UBOS with D-UBOS}

\begin{figure*}[htbp!]
    \centering
    \subfigure[]{
        \label{fig:compare_D_and_Edg_single_gate}
        \includegraphics[width=0.45\textwidth]{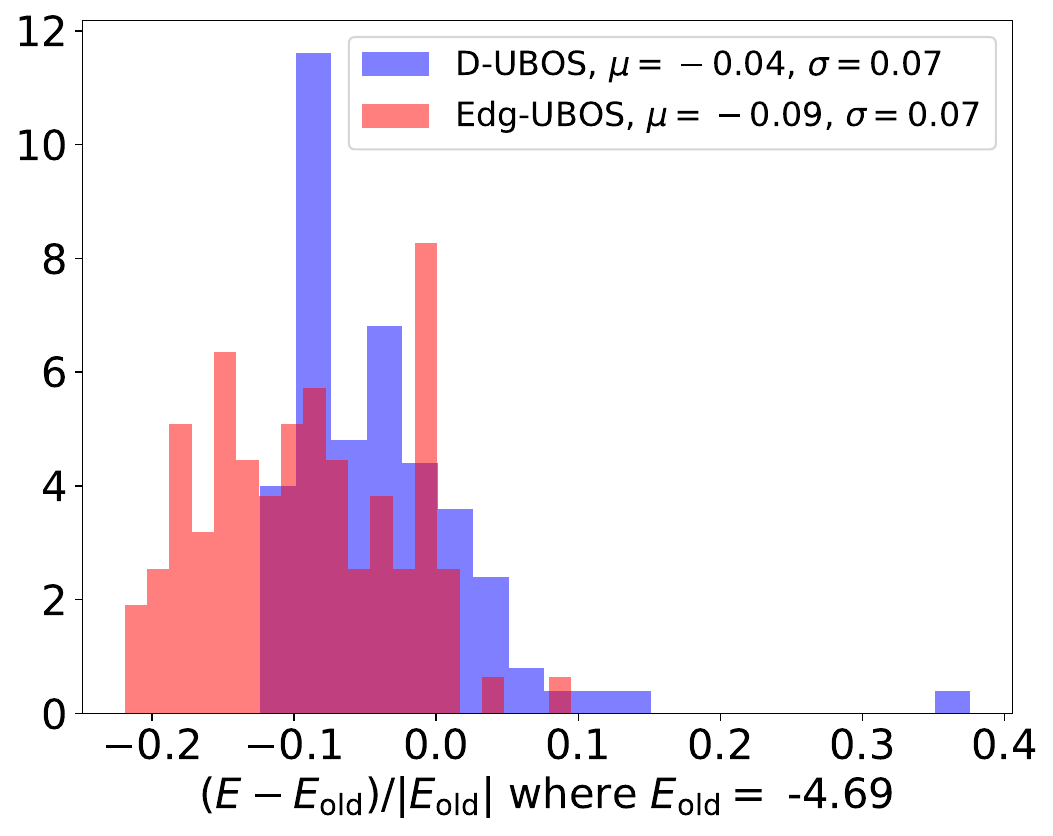}
    }
    \subfigure[]{
        \label{fig:compare_D_and_Edg_full_run_different_shots}
        \includegraphics[width=0.45\textwidth]{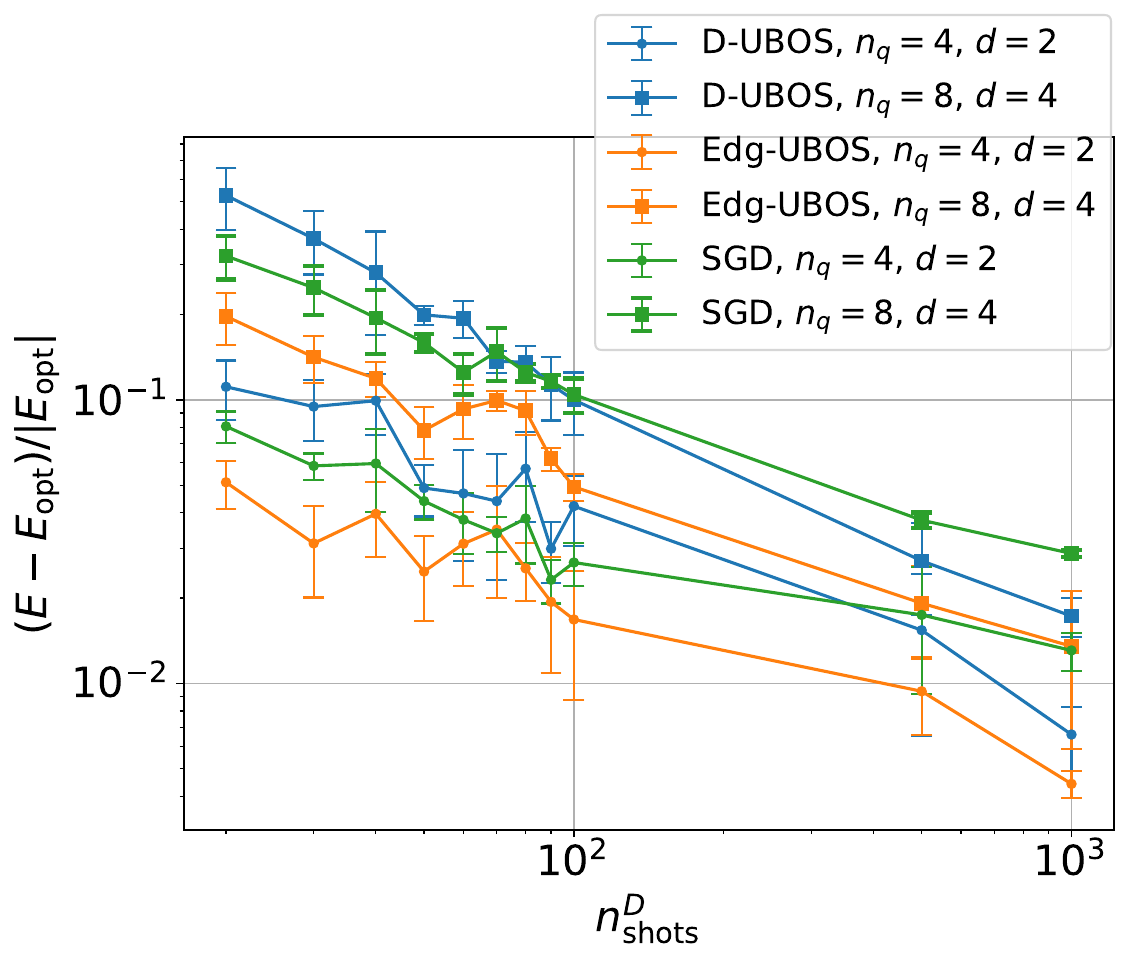}
    }
    \caption{(a) Histograms of the relative energy change from 100 steps of D-UBOS (blue) with 20 shots per circuit and Edg-UBOS (red) with 450 observations and 10 shots per circuit at the same single gate of an 8-qubit depth-4 ansatz. This corresponds to the same total number of quantum measurements for D-UBOS and Edg-UBOS. (b) Relative energy difference between 10 epochs of UBOS types (different colors) and 80 epochs of SGD on different system sizes (types of points) and the optimal VQE energy which is the minimum energy that can be obtained by the ansatz being used versus number of shots per circuit. Each point averages over the final energies of 5 independent runs with different random initial states.}
    \label{figs:alt UBOS performance for entire run}
\end{figure*}

In this paper, we use the one-dimensional quantum Heisenberg Hamiltonian with open boundary conditions
\begin{multline}
\label{eq:Hamiltonian}    
\hat{H} = -h\sum_{j=0}^{n_q}{\sigma^z_j} - J_z\sum_{j=0}^{n_q-1}{\sigma^z_j\sigma^z_{\mathrm{j+1}}}\\ - J_x\sum_{j=0}^{n_q-1}{\sigma^x_j\sigma^x_{\mathrm{j+1}}}-J_y\sum_{j=0}^{n_q-1}{\sigma^y_j\sigma^y_{\mathrm{j+1}}} 
\end{multline}
for demonstration where $n_q$ is the number of qubits and $\textit{J}_x = \textit{J}_y = \textit{J}_z = h = 1$.

To better understand the performance of UBOS with noisy expectation value measurement, we implement the relevant circuits for UBOS in Qiskit \cite{Qiskit} and perform simulations on a classical computer (without quantum hardware noise model) on a 4-site 2-layer ansatz and 8-site 4-layer ansatz. To maintain the unitarity of the two-qubit gates in optimization, the ansatz’s two-qubit unitary blocks are parameterized with the KAK decomposition \cite{KAK}. Each gate in our initial circuit is generated randomly by selecting the KAK parameters uniformly at random from $[0,\pi)$. 

To avoid ambiguity, for the rest of the paper we will use superscripts to distinguish between the number of shots per circuit for D-UBOS and for methods based on E-UBOS. For example, $n_{\textrm{shots}}^{D}$ for D-UBOS and $n_{\textrm{shots}}^{Edg}$ for Edg-UBOS.

To study the difference in optimization step quality between Edg-UBOS and D-UBOS given large shot noise, we choose the final state of a D-UBOS run after 10 epochs with 10 shots per circuit whose energy is about $60\%$ off from the optimal VQE energy (for the system size studied, the noiseless state energy plateaus before the $4^{th}$ epoch). We apply one D-UBOS step with $n_{\textrm{shots}}^{D}=20$ on the first gate. Since the result of this application is stochastic, we look at the distribution of the relative energy change, $(E-E_{\textrm{old}})/|E_{\textrm{old}}|$, over a 100 different executions of a D-UBOS step. Then we repeat this procedure using Edg-UBOS steps with $n_{\textrm{obs}}=450$ and $n_{\textrm{shots}}^{Edg}=10$ which has roughly the same total number of measurements. As shown in Fig.~\ref{fig:compare_D_and_Edg_single_gate}, we find that two distributions have roughly the same standard deviation, but the distribution for Edg-UBOS has a more negative mean value than that of D-UBOS, which indicates that one Edg-UBOS step improves the state energy more than one D-UBOS step on average. By comparing the amount of samples with positive relative change in energy, we also notice that Edg-UBOS is much less likely to find ``false positive'' gate parameters whose noiseless state energy is worse than before optimization. See Appendix~\ref{appendix:noise affecting classical optimizer} for a detailed discussion on false positive gate parameters.

We also compare the final state energies at which Edg-UBOS and D-UBOS plateau after 10 epochs (SGD after 80 epochs, see Appendix~\ref{appendix:total number of measurements}) given different shot noise. As shown in Fig.~\ref{fig:compare_D_and_Edg_full_run_different_shots}, the relative energy difference between final states of the algorithm runs and the optimal VQE energy decays approximately algebraically as $$(E-E_{\textrm{opt}})/|E_{\textrm{opt}}| \sim A \times 10^{ -\beta N_{\textrm{shots}}^D} + C$$ where A and C are algorithm and size-dependent constants with $\beta \sim 2.5\mathrm{e}{-3}$ except for one SGD ansatz (4 qubits; depth 2)  which decays with $\beta \sim 1.4\mathrm{e}{-3}$.  For the choice of measurement hyperparameters, we start with a set of $n_{\textrm{shots}}^D$ for D-UBOS and SGD and choose the combination of $(n_{\textrm{obs}}, n_{\textrm{shots}}^{Edg})$ for Edg-UBOS such that the two methods have roughly the same total number of measurements (see Appendix~\ref{appendix:total number of measurements}), prioritizing large $n_{\textrm{obs}}$. 

In the face of significant shot-noise coming from using a finite number of shots both UBOS and SGD plateau at a non-optimal VQE energy.  At a fixed number of shots, we find that D-UBOS and SGD both plateau at similar energies with SGD doing slightly better at a small number of shots and D-UBOS doing slightly better at a larger number of shots. This is consistent with what was seen in \cite{UBOS}. With the addition of Edg-UBOS, we find that the final plateaued energy is better by roughly a factor of 3 with respect to the optimal VQE energy.  

We find that the final state of an Edg-UBOS run has roughly a factor of 3 smaller relative energy error with respect to optimal VQE energy than D-UBOS for all system sizes and total number of measurements studied. 

SGD seems to outperform D-UBOS below 100 shots per circuit and is outperformed by D-UBOS as number of shots per circuit increases further, which agrees with the observations in \cite{UBOS}. For all choices of number of shots per circuit, Edg-UBOS reaches at least a factor of 2 smaller relative energy error than SGD. Note that we choose a number of epochs that is much larger than required for algorithms to plateau because energy fluctuates drastically for D-UBOS when shot noise is large and its hard to determine convergence. See Appendix.~\ref{appendix:noise affecting classical optimizer} for detail.

\begin{figure*}
    \centering
    \subfigure[]{
        \includegraphics[width=0.95\textwidth]{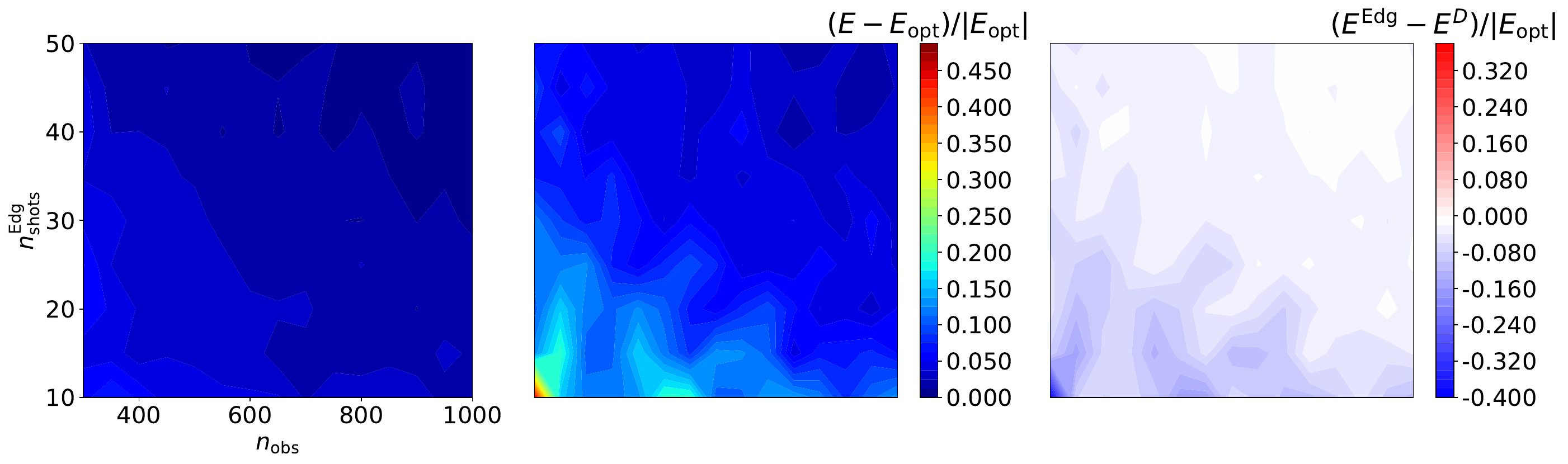}
        \label{fig:contour 4qd2}
    }
    \subfigure[]{
        \includegraphics[width=0.95\textwidth]{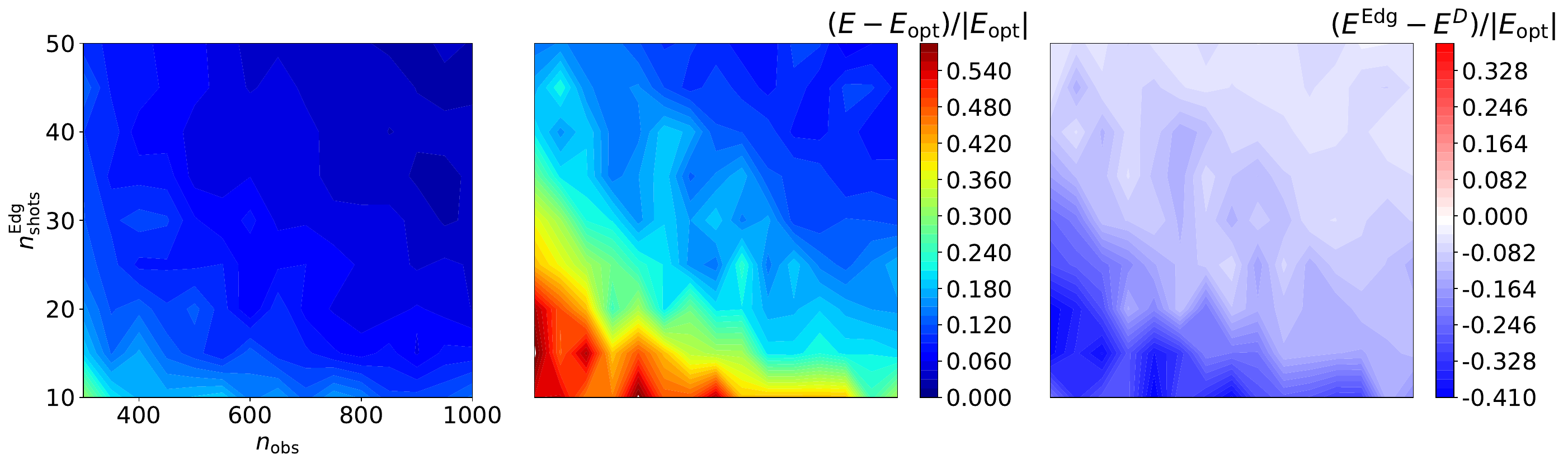}
        \label{fig:contour 8qd4}
    }
    \caption{Filled contour plots of the relative energy difference with optimal VQE energy which is the minimum energy that can be obtained by the ansatz being used, for (a) $n_q=4$ and $n_d=2$ and (b) $n_q=8$ and $n_d=4$ of Edg-UBOS (left panels), D-UBOS (middle panels), and their difference (right panels) as a function of measurement hyperparameters ($n_{\textrm{shots}}^{\textrm{Edg}}$ and $n_{\textrm{obs}}$ for Edg-UBOS, and $n_{\textrm{shots}}^{D}$ for D-UBOS with value chosen to match the total number of measurements of Edg-UBOS at each grid point). Each point averages over the final energies of 3 independent UBOS runs with different random initial states. All points in the right panels are negative, showing that Edg-UBOS reaches an energy closer to the optimal VQE energy than D-UBOS regardless of choice of measurement hyperparameters and system size.}
    \label{figs:energy contour plots}
\end{figure*}

We now consider the plateaued relative energy difference after 10 epochs of various forms of UBOS as we tune the measurement hyperparameters. As shown in Fig.~\ref{figs:energy contour plots}(left), the relative energy errors of Edg-UBOS are roughly proportional to $10^{-N_{\textrm{meas}}}$ where the total number of measurements $N_{\textrm{meas}} \propto (n_{\textrm{obs}} \times n_{\textrm{shots}}^{Edg})$ (see Appendix~\ref{appendix:total number of measurements}). For every choice of $(n_{\textrm{obs}},n_{\textrm{shots}})$ in Edg-UBOS, we can choose an identical total number of measurements in D-UBOS and again compare the relative error of the energy (see Fig.~\ref{figs:energy contour plots}(middle)) and find that for every choice of hyper-parameters, Edg-UBOS is always lower in relative energy error (on average) than D-UBOS (see Fig.~\ref{figs:energy contour plots}(right)). The advantage of Edg-UBOS becomes more significant in larger systems and when the total number of measurements are less, indicating Edg-UBOS is particularly useful when the shot noise in quantum measurements is large. See Appendix~\ref{appendix:Edg-UBOS optimal hyperparameter choice} for a more detailed discussion of optimal measurement hyperparameter choice.

\subsection{Comparing the effect of each classical technique}

\begin{figure*}
    \centering
    \subfigure[]{
        \includegraphics[width=0.45\textwidth]{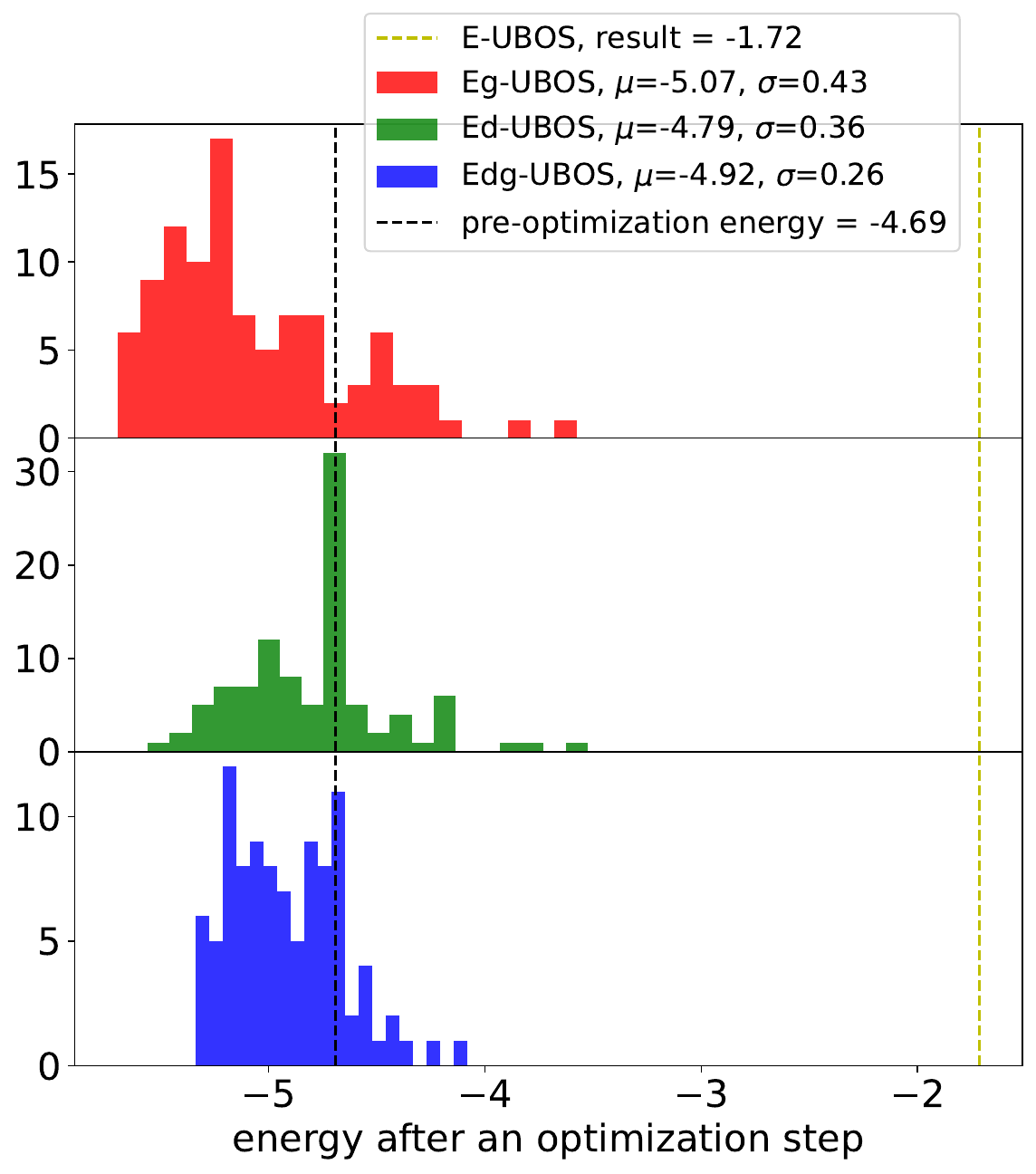}
        \label{fig:testing all methods with same training set}
    }
    \subfigure[]{
        \includegraphics[width=0.45\textwidth]{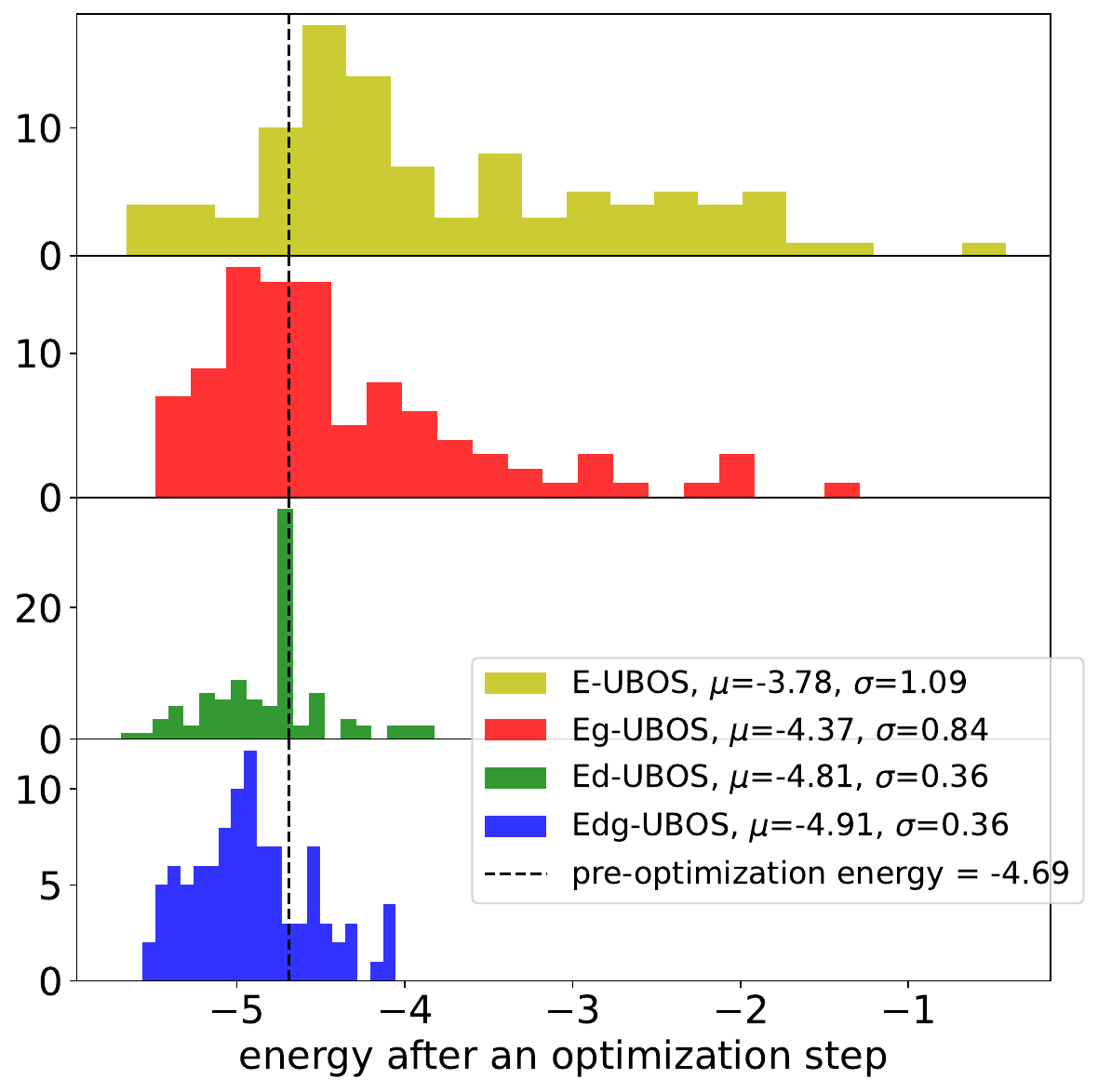}
        \label{fig:testing all methods with different training set} 
    }
   \caption{Histograms of the final energy from 100 optimizations (per approach) of the same single gate of an 8-qubit depth-4 ansatz with 10 shots per circuit. In (a) all trials use the same initial set of 300 pairs of $(t_j, E_{\textrm{noisy}})$ differing only by the randomness inherent in the techniques. In (b) a new random set of 300 parameters $t_j$ are chosen for each sample.}
    \label{figs:testing all methods}
\end{figure*}

To better understand the individual role of our two post-processing approaches, we fix a configuration of the gates and then consider the change in energy induced by the update of a single gate using these approaches. We generate a configuration of the gates by running D-UBOS for 10 epochs using only 10 shots per circuit which reaches an energy of 60\% off from the optimal VQE energy. We apply one E-UBOS optimization step (with $n_{\textrm{obs}}=300$ and $n_{\textrm{shots}}^{Edg}=10$) with different classical techniques applied (E-UBOS, Eg-UBOS, Ed-UBOS, Edg-UBOS) on the first gate, and look at the distribution of post-optimization state energy over 100 different executions for each technique.   First (see Fig.~\ref{fig:testing all methods with same training set}), we use the same fixed set of 300 observations of $(t_j, E_{\textrm{measured}})$ for all four approaches so that the 100 different executions of each approach differ due to the randomness intrinsic to each classical technique. Both classical techniques proposed in this paper as well as their combination improve the energy more often than making it worse. Moreover, all techniques and for essentially all random choices are much better than the E-UBOS step itself motivating the use of these techniques.

The improvement of the energy in the Eg-UBOS step (with GPR technique) can often be large but there is a sizeable probability of making the energy worse than the initial starting energy. The Ed-UBOS step (with DROPR technique) tests the gate parameters it's going to use with additional quantum measurements and reject the change if the energy appears to be getting worse. This means that only a small fraction of the time does the energy get worse and is responsible for the mode in the histogram at the original energy. The rest of the time the energy improves non-trivially but not as much as Eg-UBOS. The Edg-UBOS step (with both techniques) makes a good balance between the effects of both techniques. It not only makes non-trivial improvement to energy but also has strong ability to reject false-positive gate parameters after optimization.   

We further test these conclusions on the same gate (and respective configurations) by initializing 100 different initial sets of 300 observations and executing steps of \{E-UBOS, Eg-UBOS, Ed-UBOS, Edg-UBOS\} independently on each of the initial sets. As shown in Fig.~\ref{fig:testing all methods with different training set}, Eg-UBOS seems to cause more false-positive cases while having a larger chance to improve energy. Ed-UBOS detects some false-positive cases and rejects the change otherwise mainly improving the energy. Edg-UBOS (with both techniques) takes the complementary advantages of both. We also notice that the distribution of state energy after an E-UBOS step has less negative mean value than the pre-optimization energy, which indicates that an E-UBOS step worsens the energy on average. We attribute this to the fact that the minimum number of pairs is 226 and 300 pairs cannot overconstrain the system enough to reduce the large shot noise in the calculated $\tilde{H}$, which again shows the benefit of these classical post-processing techniques.

As discussed in the previous section, we think that the occurrence of false-positive cases can be greatly reduced if the number of observations is large enough to allow non-overlapping subsets of pairs being assembled in each classical techniques. 

\section{Discussion and outlook}
\label{section:discussion}

In this paper, we propose Edg-UBOS, a variant of the Unitary Block Optimization Scheme (UBOS) that is well suited for the optimization of quantum circuits on hybrid variational algorithms such as VQE.   Edg-UBOS iteratively sweeps over gates. At each step, it calculates an effective Hamiltonian \textit{$\tilde{H}$} from a system of linear equations obtained from a set of $(t_j, E_{\textrm{measured}})$ observations and then classically ﬁnds the gate parameters that minimize the energy with respect to this effective Hamiltonian while keeping the other gates temporarily fixed. Edg-UBOS implements additional classical post-processing techniques to improve the accuracy of the effective Hamiltonian calculation and the minimization of the energy. We introduced and benchmarked two schemes: data augmentation using Gaussian Process Regression and Double Robust Optimization Plus Rejection. Data augmentation only requires the original training data, making it a cost-effective approach to increasing the size and diversity of the set of observations. Meanwhile, the DROPR scheme provides a more efficient way to spend the measurement resources. The two techniques combined improves the performance of the algorithm by decreasing the final optimized error by roughly a factor of 3 largely independent of the total number of measurements made.

Edg-UBOS shares all of the standard advantages of D-UBOS including converging an order-of-magnitude faster than stochastic gradient descent (SGD), tunneling through some local minima, and having decreased sensitivity to barren plateaus \cite{UBOS}. Additionally, Edg-UBOS requires lower depths of quantum circuits and has higher resilience to shot noise. 

The total number of measurements can be further reduced by strategies such as grouping operators that can be measured jointly \cite{VQE_previous_grouping_1,VQE_previous_grouping_2,VQE_previous_grouping_3, VQE_previous_grouping_4}or by inference methods such quantum overlapping tomography \cite{VQE_previous_QOT_1}, quantum shadow tomography \cite{VQE_previous_QST_1,VQE_previous_QST_2}, and classical shadow \cite{VQE_previous_CS_1,VQE_previous_CS_2,VQE_previous_CS_3,VQE_previous_CS_4,VQE_previous_CS_5}. One can also implement adaptive number of shots per circuit so that the algorithm increases the number of shots per circuit when the energy seems to plateau, which is similar to adaptive learning rate strategy in classical machine learning.

To reach the promise of VQE we need to minimize the total number of measurements while maximizing the accuracy of the final optimization.  The development of Edg-UBOS takes an important step toward this goal and places it as one of the primary techniques for VQE on a quantum computer.  Furthermore, it also motivates an important approach toward further improving quantum algorithm through use of non-trivial classical computing resources to make the most effective use of quantum data.

\begin{acknowledgments}

We acknowledge support from the NSF Quantum Leap Challenge Institute for Hybrid Quantum Architectures and Networks (NSF Award 2016136). We acknowledge the use of scikit-learn python package \cite{scikit-learn} and IBM Quantum services \cite{ibm_osaka} for this work. This work also made use of the Illinois Campus Cluster, a computing resource that is operated by the Illinois Campus Cluster Program (ICCP) in conjunction with the National Center for Supercomputing Applications (NCSA), which is supported by funds from the University of Illinois at Urbana-Champaign.

\end{acknowledgments}

\bibliographystyle{apsrev4-1}
\bibliography{main}

\clearpage
\pagebreak

\appendix
\numberwithin{equation}{section}

\section{Parameterize the gate using KAK decomposition}
\label{appendix:KAK}
In UBOS algorithms, we parameterize the generic two-qubit unitaries by two-qubit Pauli operators \cite{Pauli_decomposition}. However, to ensure that the gate remains unitary after optimization, we also parameterize the two-qubit gate using the Cartan (KAK) decomposition for $U \in SU(4)$ \cite{KAK} as
\begin{equation}
\label{eq:KAK}
U=(A_0 \otimes A_1)(e^{-i\vec{k} \cdot \vec{\Sigma}})(B_0 \otimes B_1)
\end{equation}
where $\vec{k} \in \mathbb{R}^3$, $\vec{\Sigma} = (P^{XX},P^{YY},P^{ZZ})$, and $A_0$, $A_1$, $B_0$, and $B_1 \in SU(2)$ are generic one-qubit $U3$ gates parameterized by three real parameters as 
\begin{equation}
U3(\theta, \lambda, \phi) = 
\begin{bmatrix}
cos(\theta/2) & -e^{i\lambda}sin(\theta/2) \\
e^{i\phi}sin(\theta/2) & e^{i(\lambda+\phi)}cos(\theta/2)
\end{bmatrix}
\end{equation}

The two-qubit gate $U$ resulting from Eq.~\ref{eq:KAK} is therefore parameterized with 15 real parameters (denoted as $\theta_j$) and is unitary regardless of choice of $\theta_j$. Since the KAK decomposition we use does not have the global phase term, we cannot deterministically convert the gate parameterization from $t_j$ form to $\theta_j$ form. However, the conversion from $\theta_j$ form to $t_j$ form is deterministic since the coefficients of Pauli decomposition is unique. Therefore, to ensure the unitarity of the gate after optimization, the two-qubit unitaries of the ansatz is stored in $t_j$ form and is only converted to $\theta_j$ form before being fed to classical optimizer.  

To avoid this redundancy of using $\theta_j$, a possible approach is to perform gradient descent on a Riemannian manifold of unitary matrices \cite{Riemannian_1,Riemannian_2}.

\section{Linear least square regression for E-UBOS}
\label{appendix:E-UBOS math}

We can rewrite Eq.~\ref{eq:UBOS energy sum form} as
\begin{equation}
E(\textbf{t}_j) = \sum_{n,m=0}^{15}{t_j^{*n} \tilde{H}^{n;m} t_j^{m}} 
\label{eq:E-UBOS energy}
\end{equation}

where $m,n\in[0,15]$ are simplified notation for $(\alpha,\beta)$ and $(\alpha',\beta')$, respectively. By writing $\textit{t}_j$ and \textit{$\tilde{H}$} as their complex form, $t_j=\operatorname{Re}{[t_j]}+i\operatorname{Im}{[t_j]}$ and $\tilde{H}=\operatorname{Re}{[\tilde{H}]}+i\operatorname{Im}{[\tilde{H}]}$, we can expand Eq.~\ref{eq:E-UBOS energy} as 
\begin{multline}
E(\textbf{t}_j) = \sum_{n,m=0}^{15} \\
\operatorname{Re}{[t_j^n]}\operatorname{Re}{[\tilde{H}^{n;m}]}\operatorname{Re}{[t_j^m]} +i \operatorname{Re}{[t_j^n]}\operatorname{Im}{[\tilde{H}^{n;m}]}\operatorname{Re}{[t_j^m]} \\
+i \operatorname{Re}{[t_j^n]}\operatorname{Re}{[\tilde{H}^{n;m}]}\operatorname{Im}{[t_j^m]} - \operatorname{Re}{[t_j^n]}\operatorname{Im}{[\tilde{H}^{n;m}]}\operatorname{Im}{[t_j^m]}\\
-i \operatorname{Im}{[t_j^n]}\operatorname{Re}{[\tilde{H}^{n;m}]}\operatorname{Re}{[t_j^m]} + \operatorname{Im}{[t_j^n]}\operatorname{Im}{[\tilde{H}^{n;m}]}\operatorname{Re}{[t_j^m]} \\
+ \operatorname{Im}{[t_j^n]}\operatorname{Re}{[\tilde{H}^{n;m}]}\operatorname{Im}{[t_j^m]} +i\operatorname{Im}{[t_j^n]}\operatorname{Im}{[\tilde{H}^{n;m}]}\operatorname{Im}{[t_j^m]}
\label{eq:E-UBOS energy expanded}
\end{multline}

Notice that
\begin{multline}
\label{eq:quadratic_form}
t_j^{*n}t_j^{m} = (\operatorname{Re}{[t_j^n]}-i\operatorname{Im}{[t_j^n]})(\operatorname{Re}{[t_j^m]}+i\operatorname{Im}{[t_j^m]}) \nonumber\\
= \operatorname{Re}{[t_j^n]}\operatorname{Re}{[t_j^m]} +i\operatorname{Re}{[t_j^n]}\operatorname{Im}{[t_j^m]} \nonumber\\
-i\operatorname{Im}{[t_j^n]}\operatorname{Re}{[t_j^m]}
+ \operatorname{Im}{[t_j^n]}\operatorname{Im}{[t_j^m]} 
\end{multline}

Then by grouping terms with real and imaginary part of effective Hamiltonian in Eq.~\ref{eq:E-UBOS energy expanded}, we have
\begin{equation}
E(\textbf{t}_j) = \sum_{n,m=0}^{15} (t_j^{*n}t_j^{m}\operatorname{Re}{[\tilde{H}^{n;m}]}) +i(t_j^{*n}t_j^{m}\operatorname{Im}{[\tilde{H}^{n;m}]})
\label{eq:E-UBOS energy simplified}
\end{equation}

Since energy is a real value, Eq.~\ref{eq:E-UBOS energy simplified} can be further simplified into 
\begin{equation}   
E(\textbf{t}_j) = \sum_{n,m=0}^{15}t_{\textrm{j,R}}^{\textrm{n,m}}\operatorname{Re}{[\tilde{H}^{\textrm{n;m}}]} + t_{\textrm{j,I}}^{\textrm{n,m}}\operatorname{Im}{[\tilde{H}^{\textrm{n;m}}]} 
\label{eq:E-UBOS energy final}
\end{equation} 
where 
\begin{align}
t_{\textrm{j,R}}^{\textrm{n,m}} &\equiv \operatorname{Re}{[t_j^{n} (t_j^{m})^*]} \nonumber \\
t_{\textrm{j,I}}^{\textrm{n,m}} &\equiv \operatorname{Im}{[t_j^{n} (t_j^{m})^*]} \nonumber
\end{align}
are quadratic forms of the $\textit{t}_j$ components 

For linear least square regression, we can write Eq.~\ref{eq:E-UBOS energy final} as 
\begin{equation}   
E(\textbf{t}_j) = \langle\operatorname{Re}{[t_{\textrm{j,Q}}]}, \operatorname{Re}{[\tilde{H}]}\rangle_F + \langle\operatorname{Im}{[t_{\textrm{j,Q}}]}, \operatorname{Im}{[\tilde{H}]}\rangle_F
\label{eq:E-UBOS energy linear regression form}
\end{equation}

where $t_{\textrm{j,Q}}=t_j \otimes t_j^*$ is the outer product of $t_j$ with its complex conjugate and $\langle A,B\rangle_F$ is the Frobenius inner product of two matrices $A$ and $B$

\section{Measurement circuit for different types of UBOS and SGD}
\label{appendix:UBOS measurement circuits}

\begin{figure}[htbp!]
    \centering
    \Qcircuit@C=0.7em@R=.7em{
        &\mbox{\space\space\space\space\space\space\space\space\space\space\space\space\space\space\space\space\space\space\space\space\space\space\space\space\space\space\space\space\space\space\space\space\space\space\space\space\space\space\space\space\space\space\space\space\space\space\space\space\space\space\space \ket{\psi} } \\
        &\mbox{\space}\\
        \push{\rule{2em}{0em}} &\lstick{\ket{0}} &\qw &\multigate{1}{U_1} &\qw &\qw &\qw &\qw &\multigate{1}{U_6} &\qw &\qw &\qw &\qw &\qw &\qw\\  
        \push{\rule{2em}{0em}} &\lstick{\ket{0}} &\qw &\ghost{U_1} &\qw &\multigate{1}{U_4} &\qw &\qw &\ghost{U_6} &\qw &\multigate{1}{U_9} &\qw &\qw &\qw &\qw\\ 
        \push{\rule{2em}{0em}} &\lstick{\ket{0}} &\qw &\multigate{1}{U_2} &\qw &\ghost{U_4} &\qw &\qw &\multigate{1}{U_7} &\qw &\ghost{U_9} &\qw &\multigate{1}{r_{\hat{h}}} &\qw &\meter &\qw\\  
        \push{\rule{2em}{0em}} &\lstick{\ket{0}} &\qw &\ghost{U_2} &\qw &\multigate{1}{U_5} &\qw &\qw &\ghost{U_7} &\qw &\multigate{1}{U_{10}} &\qw &\ghost{r_{\hat{h}}} &\qw &\meter &\qw \\
        \push{\rule{2em}{0em}} &\lstick{\ket{0}} &\qw &\multigate{1}{U_3} &\qw &\ghost{U_5} &\qw &\qw &\multigate{1}{U_8} &\qw &\ghost{U_{10}} &\qw &\qw &\qw &\qw\\  
        \push{\rule{2em}{0em}} &\lstick{\ket{0}} &\qw &\ghost{U_3} &\qw &\qw &\qw &\qw &\ghost{U_8} &\qw &\qw &\qw &\qw &\qw &\qw \gategroup{3}{4}{8}{12}{0.6em}{--}
        }
    \caption{Quantum circuit for E-UBOS measurement. The circuit shown is an example of measuring the expectation value of a two-qubit operator $\hat{h}$ which is a component of the Hamiltonian acting on the 3rd and 4th qubit using a 6-qubit depth-4 ansatz. The $r_{\hat{h}}$ gate changes the basis to the eigenbasis of $\hat{h}$. For example, $r_{\hat{h}}$ gate being two Hadamard gates changes the basis to $\{+, -\}$ basis for measuring $\hat{h} = P^{XX}$.}
    \label{fig:E-UBOS measurement circuit}
\end{figure}
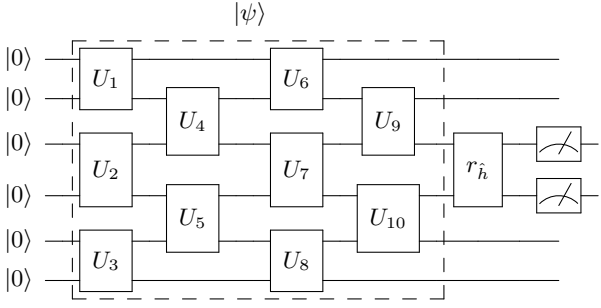

The energy of a quantum state can be found as $E=\sum_i{\bra{\psi}\hat{h}_i\ket{\psi}}$ where $\hat{h}_i$ are components of the Hamiltonian. In E-UBOS, we obtain the state energy by measuring all Hamiltonian components with circuit shown in Fig.~\ref{fig:E-UBOS measurement circuit}. Since the Hamiltonian studied in this work contains only local $\{Z, XX, YY, ZZ\}$ operators, we obtain the expectation values in the following way:

\begin{align}
    \bra{\psi}\hat{Z}\ket{\psi} &\approx P(0)-P(1) \nonumber \\
    \bra{\psi}\widehat{ZZ}\ket{\psi} &\approx P(0,0)-P(0,1)-P(1,0)+P(1,1) \nonumber\\
    \bra{\psi}\widehat{XX}\ket{\psi} &\approx 2P(+,+)+2P(-,-)-1 \nonumber\\
    \bra{\psi}\widehat{YY}\ket{\psi} &\approx 2P(+i,+i)+2P(-i,-i)-1 \nonumber\\
\end{align}
where $P(\cdot)$ is the relative frequency of measuring corresponding states from sampling, and these expectation values become exact with an infinite number of shots. By linear combination with their corresponding coefficients (which are all equal to 1 in this paper), we obtain the energy of the state.

The Hadamard test circuit for $\tilde{H}$ matrix element measurement in D-UBOS and SGD is shown in Fig.~\ref{fig:UBOS circuit diagram}. See \cite{UBOS} for a detailed guide on using this measurement circuit.

\begin{figure*}[htbp!]
    \centering
    \Qcircuit@C=0.7em@R=.7em{
        &\mbox{\space\space\space\space\space\space\space\space\space\space\space\space\space\space\space\space\space\space\space\space\space\space\space\space\space\space\space\space\space\space\space\space\space\space\space\space\space\space\space\space\space\space\space\space\space\space\space\space\space\space\space\space\space\space\space\space\space\space\space\space\space\space\space\space\space\space\space\space\space\space\space\space\space\space\space\space\space\space\space\space\space\space\space\space\space\space\space\space\space\space\space\space\space\space\space\space\space\space\space\space\space\space\space\space\space\space\space\space\space\space\space\space\space\space\space\space\space\space \ket{\psi_{\mathrm{j=5}}^{\mathrm{\alpha\beta}}} \space\space\space\space\space\space\space\space\space\space\space\space\space\space\space\space\space\space\space\space\space\space\space\space\space\space\space\space\space\space\space\space\space\space\space\space\space\space\space\space \bra{\psi_{\mathrm{j=5}}^{\mathrm{\alpha'\beta'}}} up to j=5 } \\
        &\mbox{\space}\\
        \push{\rule{2em}{0em}} &\lstick{\ket{0}} &\qw &\multigate{1}{U_1} &\qw &\qw &\qw &\qw &\multigate{1}{U_6} &\qw &\qw &\qw &\qw &\qw &\qw &\qw &\multigate{1}{U_6^\dag} &\qw &\qw &\qw &\qw \\
        \push{\rule{2em}{0em}} &\lstick{\ket{0}} &\qw &\ghost{U_1} &\qw &\multigate{1}{U_4} &\qw &\qw &\ghost{U_6} &\qw &\multigate{1}{U_9} &\qw &\qw &\qw &\multigate{1}{U_9^\dag} &\qw &\ghost{U_6^\dag} &\qw &\qw &\qw &\qw \\
        \push{\rule{2em}{0em}} &\lstick{\ket{0}} &\qw &\multigate{1}{U_2} &\qw &\ghost{U_4} &\qw &\qw &\multigate{1}{U_7} &\qw &\ghost{U_9} &\qw &\multigate{1}{\hat{h}} &\qw &\ghost{U_8^\dag} &\qw &\multigate{1}{U_7^\dag} &\qw &\qw &\qw &\qw \\
        \push{\rule{2em}{0em}} &\lstick{\ket{0}} &\qw &\ghost{U_2} &\qw &\multigate{1}{P_5^{\alpha\beta}} &\qw &\qw &\ghost{U_7} &\qw &\multigate{1}{U_{10}} &\qw &\ghost{\hat{h}} &\qw &\multigate{1}{U_{10}^\dag} &\qw &\ghost{U_7^\dag} &\qw &\multigate{1}{P_5^{\alpha^\prime\beta^\prime}} &\qw &\qw \\
        \push{\rule{2em}{0em}} &\lstick{\ket{0}} &\qw &\multigate{1}{U_3} &\qw &\ghost{P_5^{\alpha\beta}} &\qw &\qw &\multigate{1}{U_8} &\qw &\ghost{U_{10}} &\qw &\qw &\qw &\ghost{U_{10}^\dag} &\qw &\multigate{1}{U_8^\dag} &\qw &\ghost{P_5^{\alpha^\prime\beta^\prime}} &\qw &\qw \\
        \push{\rule{2em}{0em}} &\lstick{\ket{0}} &\qw &\ghost{U_3} &\qw &\qw &\qw &\qw &\ghost{U_8} &\qw &\qw &\qw &\qw &\qw &\qw &\qw &\ghost{U_8^\dag} &\qw &\qw &\qw &\qw \\
        \push{\rule{2em}{0em}} &\lstick{ancilla} &\gate{H} &\qw &\qw &\ctrl{-2} &\qw &\qw &\qw &\qw &\qw &\qw &\ctrl{-3} &\qw &\qw &\qw &\qw &\qw &\ctrl{-2} &\qw &\gate{S^b} &\qw &\gate{H} &\qw &\rstick{-\expval{Z}} \gategroup{3}{4}{8}{12}{0.6em}{--} \gategroup{3}{14}{8}{20}{1.2em}{--}
        }
    \caption{Quantum circuit for measuring $\tilde{H}$ matrix elements in D-UBOS and estimating gradient in SGD. The circuit shown is an example for measuring $\tilde{H}^{\mathrm{\alpha'\beta';\alpha\beta}}$ at $j=5$th gate using a 6-qubit, depth-4 ansatz. The unitary operator $\hat{h}$ is one of the Hamiltonian components. The $j=5$th gate of \ket{\psi_{\mathrm{j=5}}^{\mathrm{\alpha\beta}}}, \bra{\psi_{\mathrm{j=5}}^{\mathrm{\alpha'\beta'}}}, and $\hat{h}$ are controlled by the ancilla qubit. The adjoint of the gates are applied to the circuit up to $j=5$th gate because the rest of the gates will not affect the measurement outcome. When the boolean $\textit{b}=0$ $(b=1)$, the real (imaginary) part of the expectation value is estimated by the negative of the expectation value of Pauli-Z operator. }
    \label{fig:UBOS circuit diagram}
\end{figure*}

\section{Calculating total number of measurements}
\label{appendix:total number of measurements}

The total number of measurements is defined as 
\begin{equation}
N_{\textrm{meas}} = N_{\textrm{step}} \times N_{\textrm{element}} \times N_{\textrm{operator}} \times n_{\textrm{shots}} 
\end{equation}
where $N_{\textrm{step}}=N_{\textrm{epoch}} \times N_{\textrm{gate}}$ is the number of optimization steps, the number of gates in the ansatz $N_{\textrm{gate}}$ is a constant determined by the number of qubits $n_q$ and the circuit depth $d$, the number of training epochs $N_{\textrm{epoch}}$ is a algorithm hyperparameter, $N_{\textrm{element}}$ is the number of matrix elements or observations to obtain the effective Hamiltonian, and $N_{\textrm{operator}}$ is the number of unique operators in the Hamiltonian. 

The Hamiltonian studied contains only local $\{Z, XX, YY, ZZ\}$ operators. 
For an ansatz of $n_q$ qubits, we need to measure $n_q$ unique $Z$ operators acting on different qubits. For the three kinds of two-qubit operators, due to the open boundary condition of the ansatz, there exists $(n_q-1)$ unique operators of each kind. Therefore, the total number of unique operators in the Hamiltonian is $4 n_q-3$. 

The total number of measurements for D-UBOS and Edg-UBOS are calculated as
\begin{align}
N_{\textrm{meas}}^{SGD} &= N_{\textrm{step}} \times 32 \times (4n_q-3) \times n_{\textrm{shots}}^{D} \label{eq:SGD N measurement} \\
N_{\textrm{meas}}^{D} &= N_{\textrm{step}} \times 256 \times (4n_q-3) \times n_{\textrm{shots}}^{D} \label{eq:UBOS N measurement} \\
N_{\textrm{meas}}^{Edg} &= N_{\textrm{step}} \times (n_{\textrm{obs}}+60) \times (4n_q-3) \times n_{\textrm{shots}}^{Edg} \label{eq:alt-UBOS N measurement}
\end{align}
where 32 is the unique real parameters of the gradient estimator for SGD, 256 is the unique real parameters of the effective Hamiltonian, and 60 is the empirically-chosen number of observations for testing the contender gate parameters in DROPR scheme. For simplicity we keep the number of shots per circuit the same for D-UBOS and SGD. Note that $n_{\textrm{obs}}$ has a minimum value of 226.

\section{How noise makes the optimization plateau}
\label{appendix:noise affecting classical optimizer}

\begin{figure*}
    \centering
    \subfigure[]{
        \includegraphics[width=0.45\textwidth]{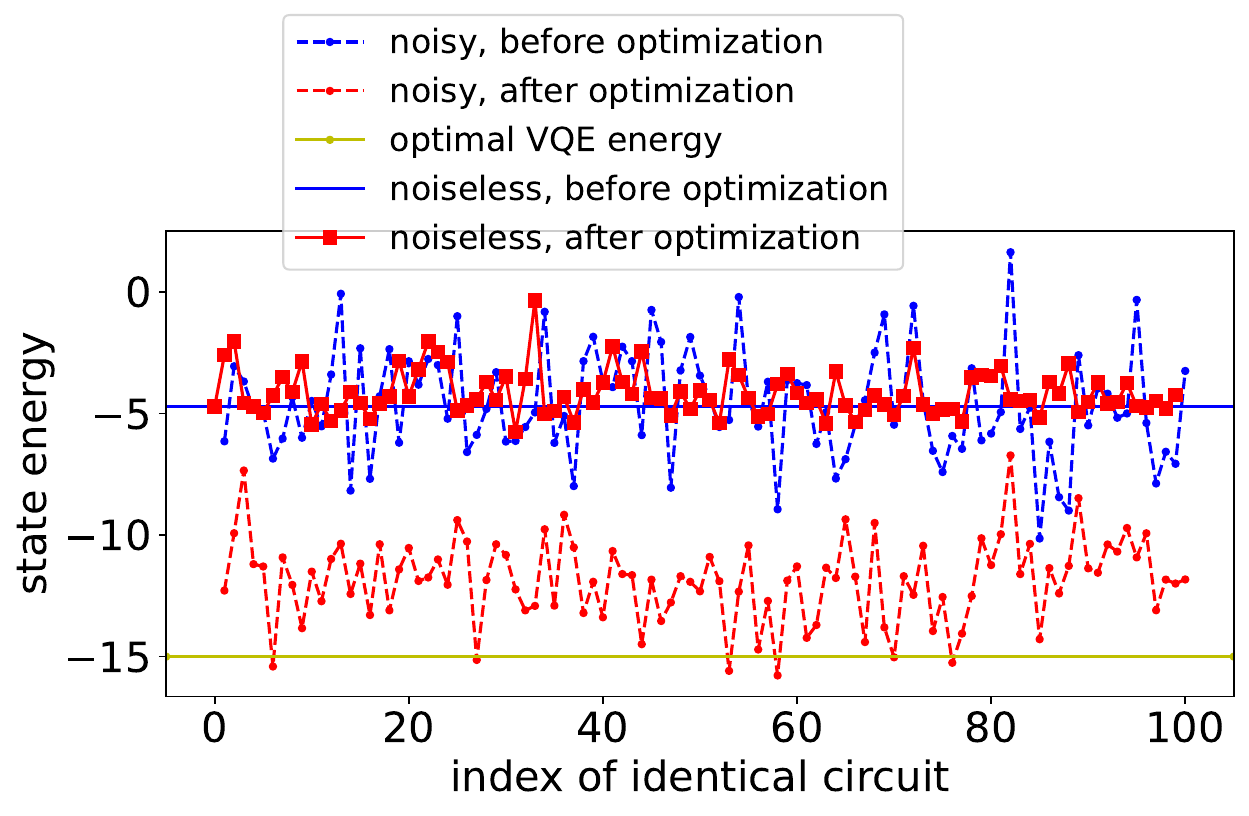}
        \label{figs:D-UBOS false positive}
    }
    \subfigure[]{
        \includegraphics[width=0.45\textwidth]{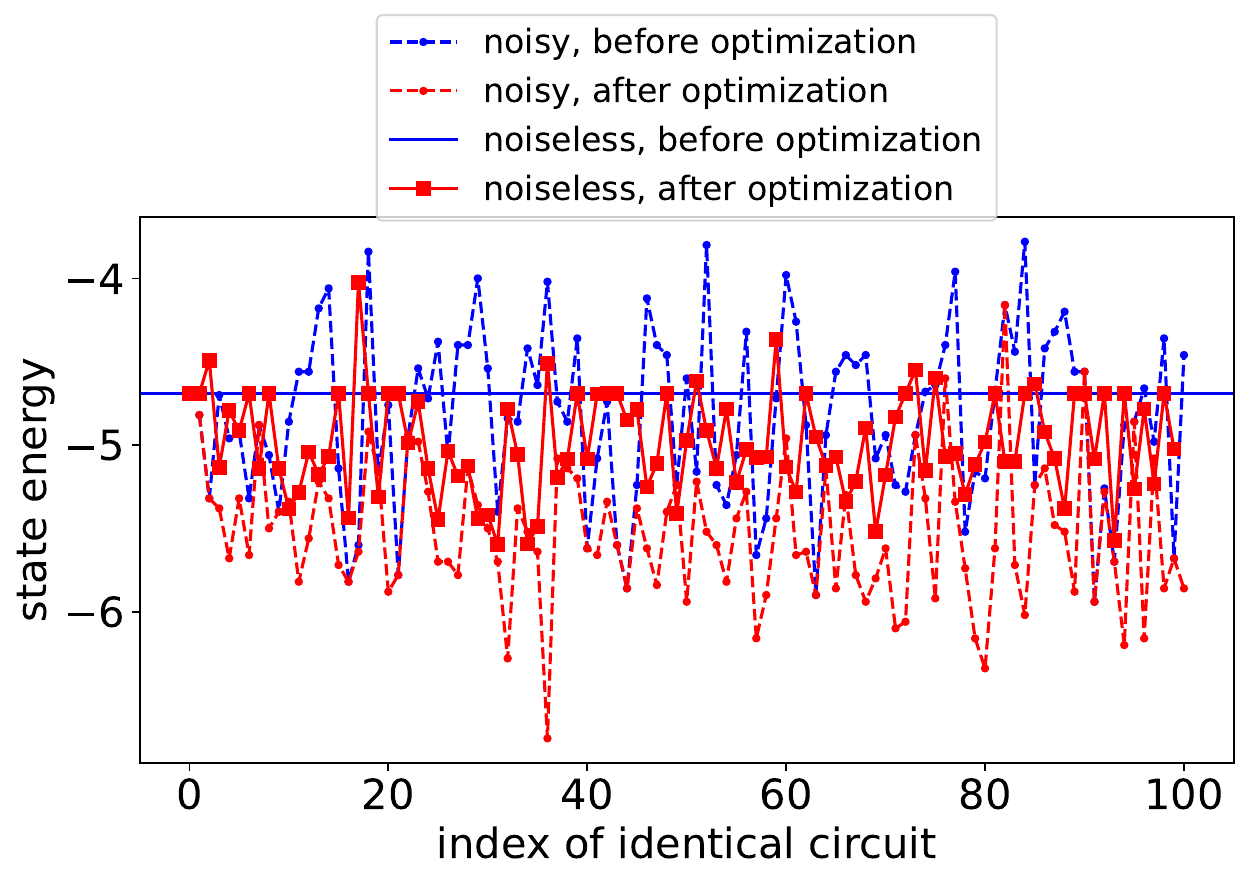}
        \label{fig:Edg-UBOS false positive} 
    }
    \caption{The change in noisy state energy and noiseless state energy after (a) a D-UBOS and (b) an Edg-UBOS optimization step on the same gate of 100 identical circuits using ansatz of 8-qubit depth-4. The noisy state energy is always improved by the classical optimizer using the noisy $\tilde{H}$. For D-UBOS, the noiseless energy worsens after optimization on average; for Edg-UBOS, the noiseless energy improves on average (the optimal VQE energy is -15). }
    \label{fig:stupid classical optimizer}
\end{figure*}

To understand how noisy measurement outcomes lead to energy plateauing above the optimal energy, we consider the change in noisy state energy and in noiseless state energy after one optimization step. We choose the state of a D-UBOS run after 10 epochs with 10 shots per circuit whose energy is about 60\% off from the optimal VQE energy. We apply one D-UBOS step with $n_{\textrm{shots}}^{D}=20$ on the same gate of 100 identical state. As shown in Fig.~\ref{figs:D-UBOS false positive}, for a D-UBOS step, the classical optimizer always improves the noisy energy calculated with $\tilde{H}$, but the noiseless energy gets worse in most cases and is very different from the noisy energy, which we refer to as a "false-positive case". The excessive noise in measured $\tilde{H}$ makes it possible for classical optimizer to find an unphysical noisy state energy below the optimal VQE energy of the ansatz. Then we repeat this procedure using Edg-UBOS steps with $n_{\textrm{obs}}=450$ and $n_{\textrm{shots}}^{Edg}=10$ which has roughly the same total number of measurements (see Fig.~\ref{fig:Edg-UBOS false positive}). We find that even though the improvement in noisy energy by an Edg-UBOS step is much smaller than that by a D-UBOS step, the noiseless energy is improved in most cases, which indicates that Edg-UBOS can effectively suppress the noise in the effective Hamiltonian (due to noisy quantum measurements) and make non-trivial improvements to the noiseless energy. Besides, the noiseless energy after an Edg-UBOS step is much closer to the noisy energy, showing the algorithm's high accuracy in state energy estimation, which is very important for convergence detection towards the end of its run.

\begin{figure}[htbp!]
    \centering
    \includegraphics[width=0.45\textwidth]{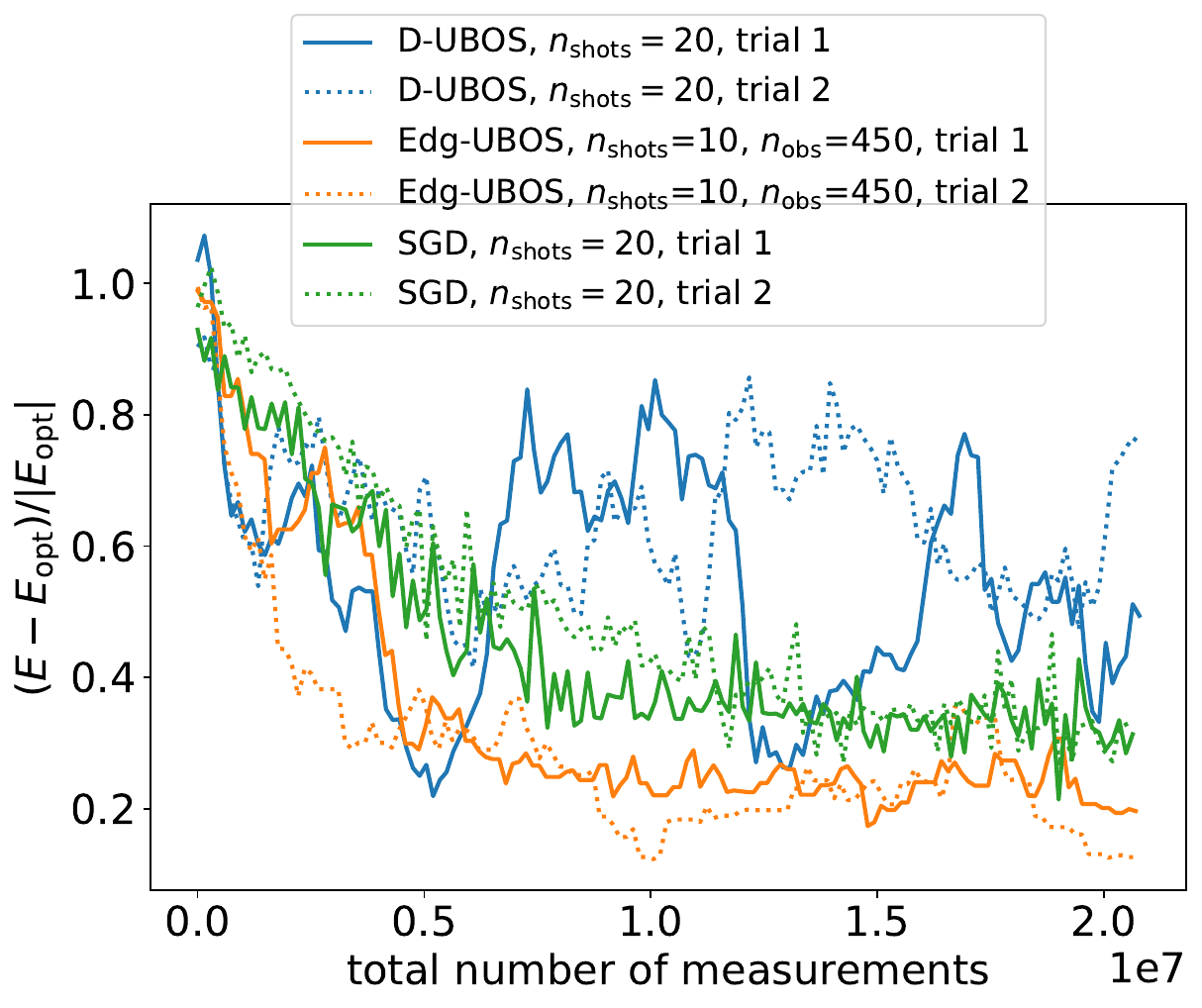}
    \caption{Relative energy error for $n_q=8$ and $n_d=4$ as a function of total number of measurements during two different optimization runs of D-UBOS (blue; 20 shots per circuit), Edg-UBOS (orange; 10 shots per circuit, 450 observations), and SGD (green; 20 shots per circuit). which correspond to the same number of measurements per UBOS step).}
    \label{fig:change of relative energy error during full run}
\end{figure}

Then we independently execute two D-UBOS runs with 20 shots per circuit, two Edg-UBOS runs with 450 observations and 10 shots per circuit, and two SGD runs on different random initial states. We look at the change in relative energy error throughout the full runs (see Fig.~\ref{fig:change of relative energy error during full run}). We find that Edg-UBOS always plateaus at an energy level much better than D-UBOS and SGD. Moreover, even though the relative energy error can get worse after one optimization step due to noise in quantum measurement outcomes, the scale of such setback in Edg-UBOS is much smaller than in D-UBOS and is comparable to it in SGD, which demonstrates that Edg-UBOS is much favorable for variational algorithms.

\section{GPR parameters}
\label{appendix:GPR parameters}

The GPR scheme has several hyperparameters including the number and size of subsets, the choice of kernel function, and the number of artificial data generated with each model. In this paper, we empirically choose to make 60 subsets with size equal to 60\% of the initial set size. We use the Radial Basis Function (RBF) kernel with length scale of 1 for the Gaussian Process Regressor, which is the common default. The number of artificial data generated per model is 2\% of the initial set size. 

\begin{figure}[htbp!]
    \centering
    \includegraphics[width=0.45\textwidth]{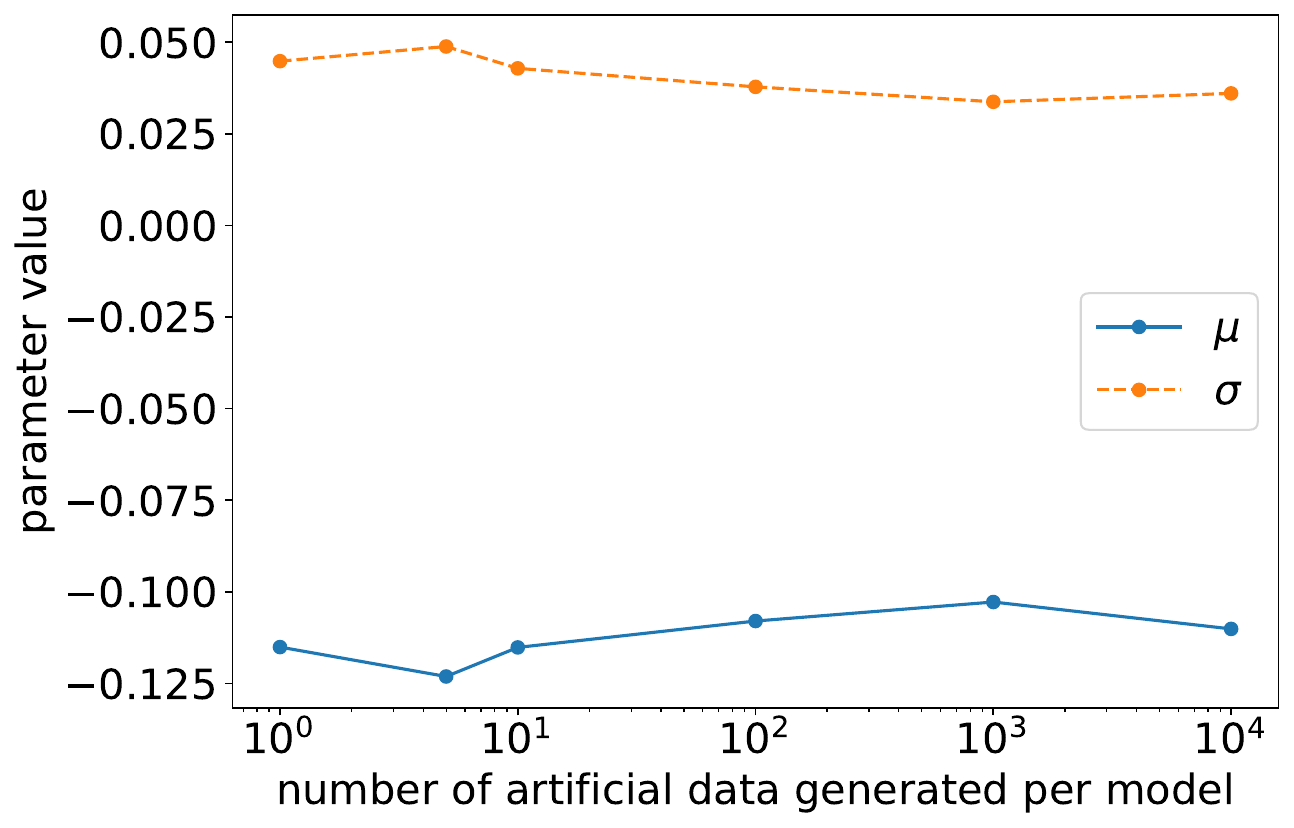}
    \caption{Mean (blue solid line) and standard deviation (orange dashed line) of distributions of post-optimization state energy from 100 different optimization steps on the first gate of identical state as a function of numbers of artificial data generated per Gaussian Process model.}
    \label{fig:GPR n_fake_data}
\end{figure}

In principle, one can generate arbitrarily large amount of artificial data at the cost of classical computing resources. To understand the relationship between the amount of artificial data and the effect of the GPR scheme, we fix a configuration of the gates and then consider the change in energy induced by the update of a single gate using Eg-UBOS with different number of artificial data generated per model. Again we choose the state of a D-UBOS run after 10 epochs with 10 shots per circuit whose energy is about 60\% off from the optimal VQE energy. We apply one Eg-UBOS optimization step (with $n_{\textrm{obs}}=450$ and $n_{\textrm{shots}}^{Edg}=10$) with different number of artificial data generated per model on the first gate, and look at the mean and standard deviation of the distributions of post-optimization state energy over 100 different executions for each hyperparameter choice. As shown in Fig.~\ref{fig:GPR n_fake_data}, both the mean and the standard deviation of the distribution don't change much as the number of artificial data generated per model increases. Since the number of subsets is 60, generating 10 artificial pairs per model means 600 artificial pairs which is already larger than the amount of pairs in the initial set which is 450. Therefore, it is unlikely to further increase the diversity and comprehensiveness of the augmented data set by further generation of artificial data, leading to similar mean and standard deviation of the distribution of post-optimization state energy for all choices of number of artificial data generated per model.

\begin{figure}[htbp!]
    \centering
    \includegraphics[width=0.45\textwidth]{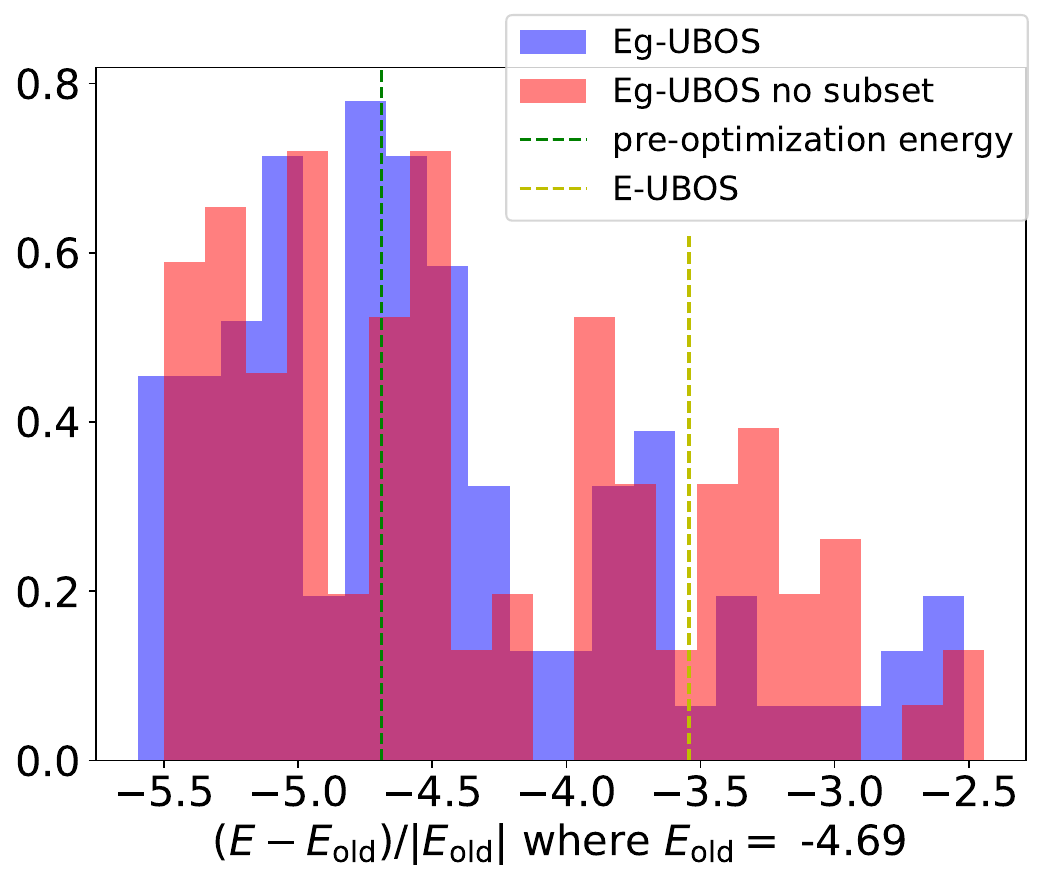}
    \caption{Histograms of the final energy from 100 optimizations (per approach) of the same single gate of an 8-qubit depth-4 ansatz with 10 shots per circuit. All trials use the same initial set of 300 pairs of $(t_j, E_{\textrm{noisy}})$ differing only by the randomness inherent in the techniques. }
    \label{fig:GPR subsetting}
\end{figure}

We compare the normal GPR scheme to the GPR scheme training only one GPR model with the entire training set for artificial data generation. As shown in Fig.~\ref{fig:GPR subsetting}, when some measured $(t, E)$ pairs in the initial set are so noisy that one E-UBOS update worsens the state energy, dividing the initial set into overlapping subsets and generating artificial data from these subsets causes significantly fewer cases of worsening the energy than generating artificial data from the entire initial set.

\section{DROPR energy measurement for contender gate parameters}
\label{appendix:DROPR energy measurement}

\begin{figure}[htbp!]
    \centering
    \includegraphics[width=0.45\textwidth]{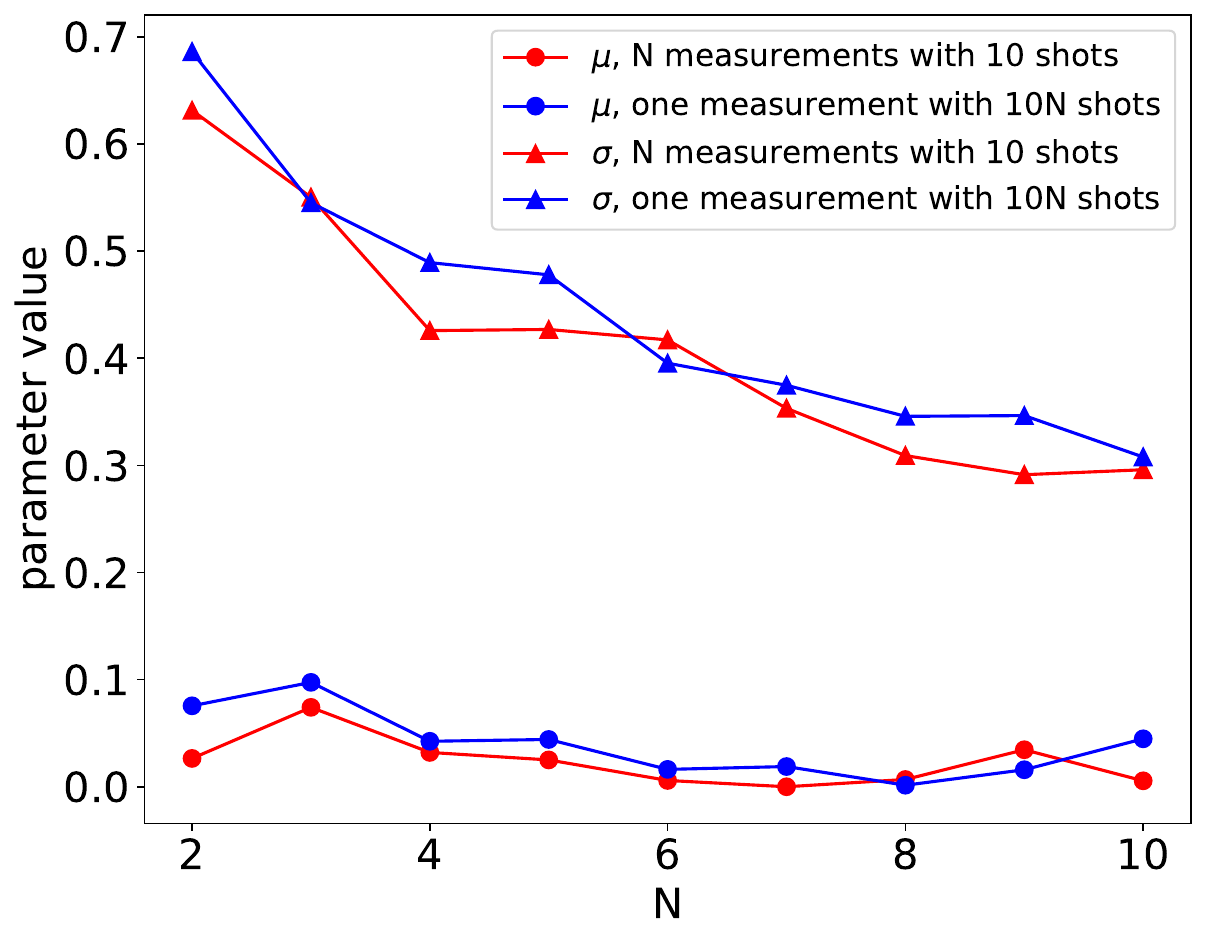}
    \caption{Mean (circle marker) and standard deviation (triangle marker) of distribution of state energy errors measured one time with $10N$ (blue) shots per circuit and measured $N$ time with 10 shots (red) as a function of $N$.}
    \label{fig:accurate measurement}
\end{figure}

In the DROPR scheme, we find the energy of the state assembled with each contender gate parameters $t_j$ through additional quantum measurements. Even though we spend more shots per circuit for these measurements, the outcome is still going to be noisy. An alternative approach is to measure the same state several times, each with the same shots per circuit as in measuring observations ($n_{\textrm{shots}}^{Edg}$), and averaging the obtained energies. To spend the measurement resources more efficiently, we compare the accuracy of measured energy obtained by measuring the state one time with $10N$ shots per circuit to measuring the identical state $N$ times with 10 shots per circuit and averaging the measured energies. We choose a random state and measure its energy 100 times independently with each of these two methods, and repeat with different values of $N$. We look at the mean and standard deviation of the distributions of the energy error (see Fig.~\ref{fig:accurate measurement}). We find that, when $N$ is small such that the $(10N)$-shots measurement outcome is still very noisy, averaging over many noisy measured energies is slightly more accurate than one less noisy measured energies. Some interesting open questions include that whether this conclusion holds as $N$ further increases and that if there exists a deterministic optimal shots per circuit for each measurement instead of 10 which is empirically chosen.

\section{Edg-UBOS optimal hyperparameter choice}
\label{appendix:Edg-UBOS optimal hyperparameter choice}

\begin{figure*}
    \centering
    \subfigure[]{
        \includegraphics[width=0.45\textwidth]{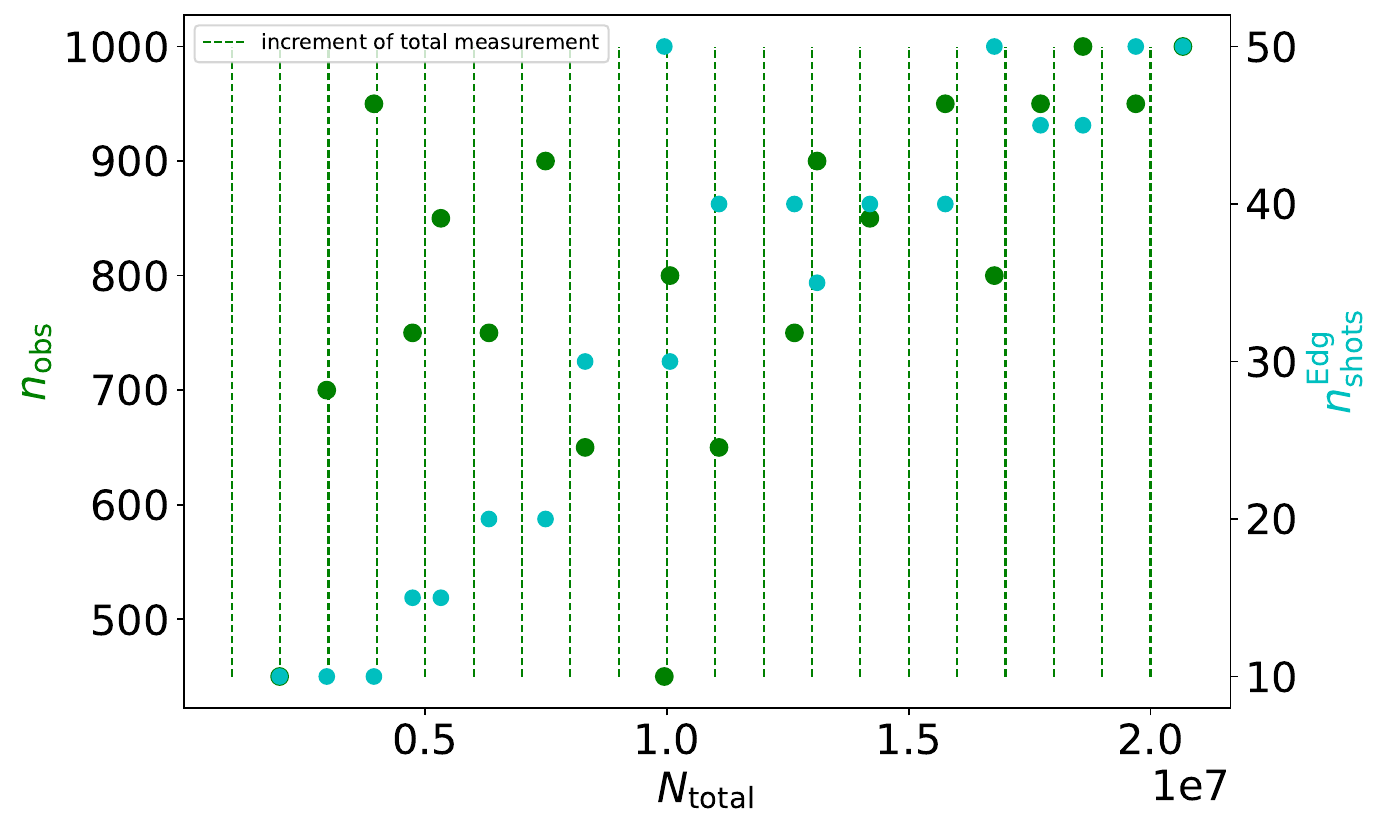}
        \label{fig:optimal parameter choice 4qd2}
    }
    \subfigure[]{
        \includegraphics[width=0.45\textwidth]{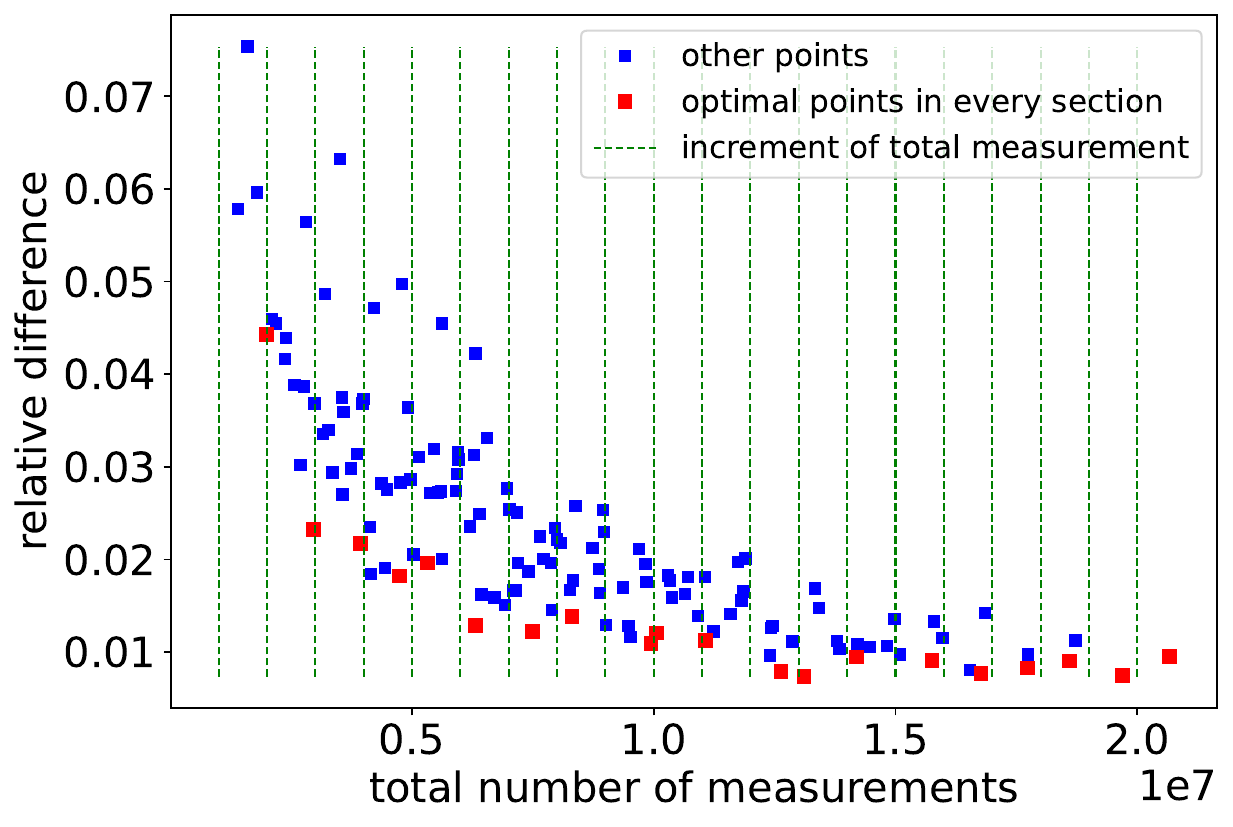}
        \label{fig:optimal parameter wins 4qd2}
    }
    \subfigure[]{
        \includegraphics[width=0.45\textwidth]{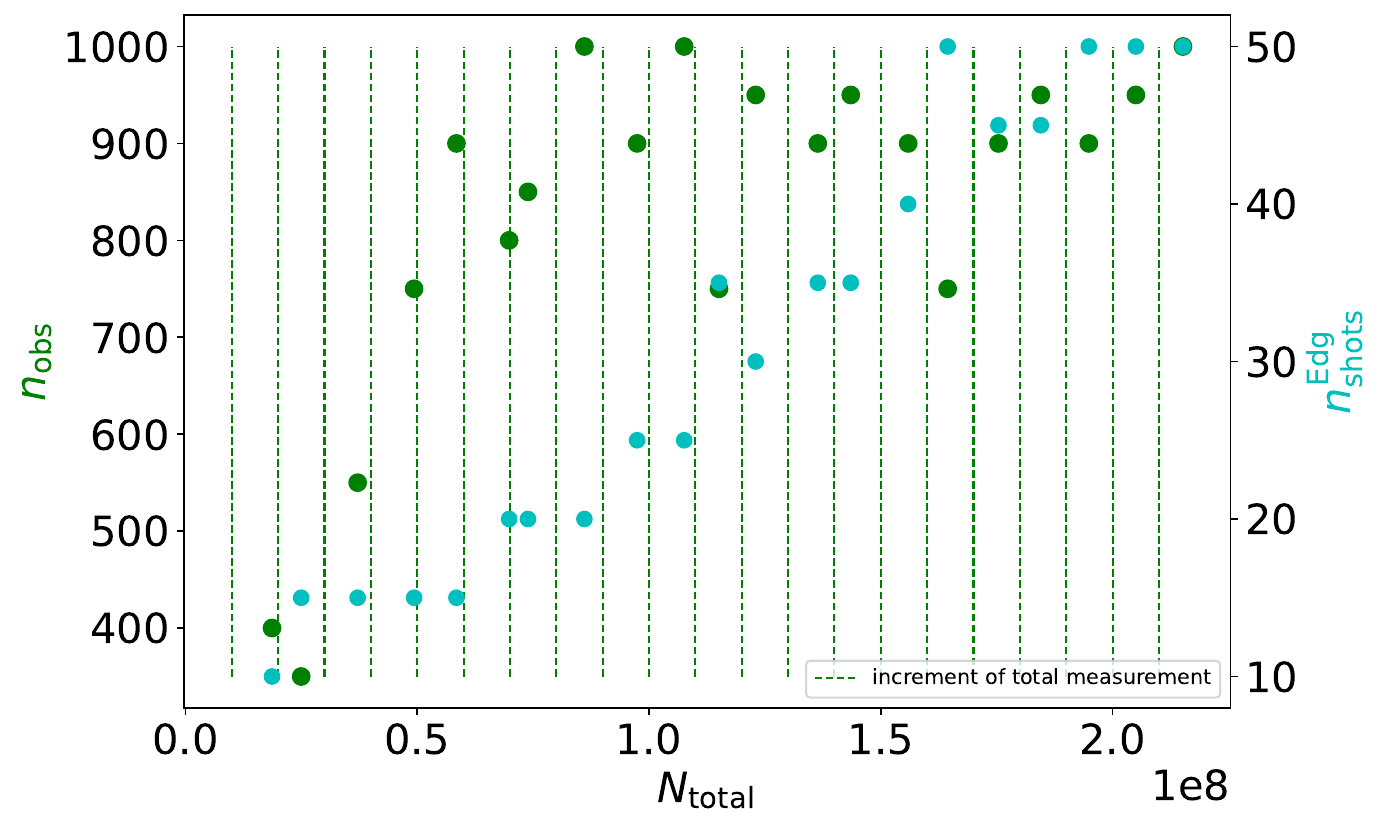}
        \label{fig:optimal parameter choice 8qd4} 
    }
    \subfigure[]{
        \includegraphics[width=0.45\textwidth]{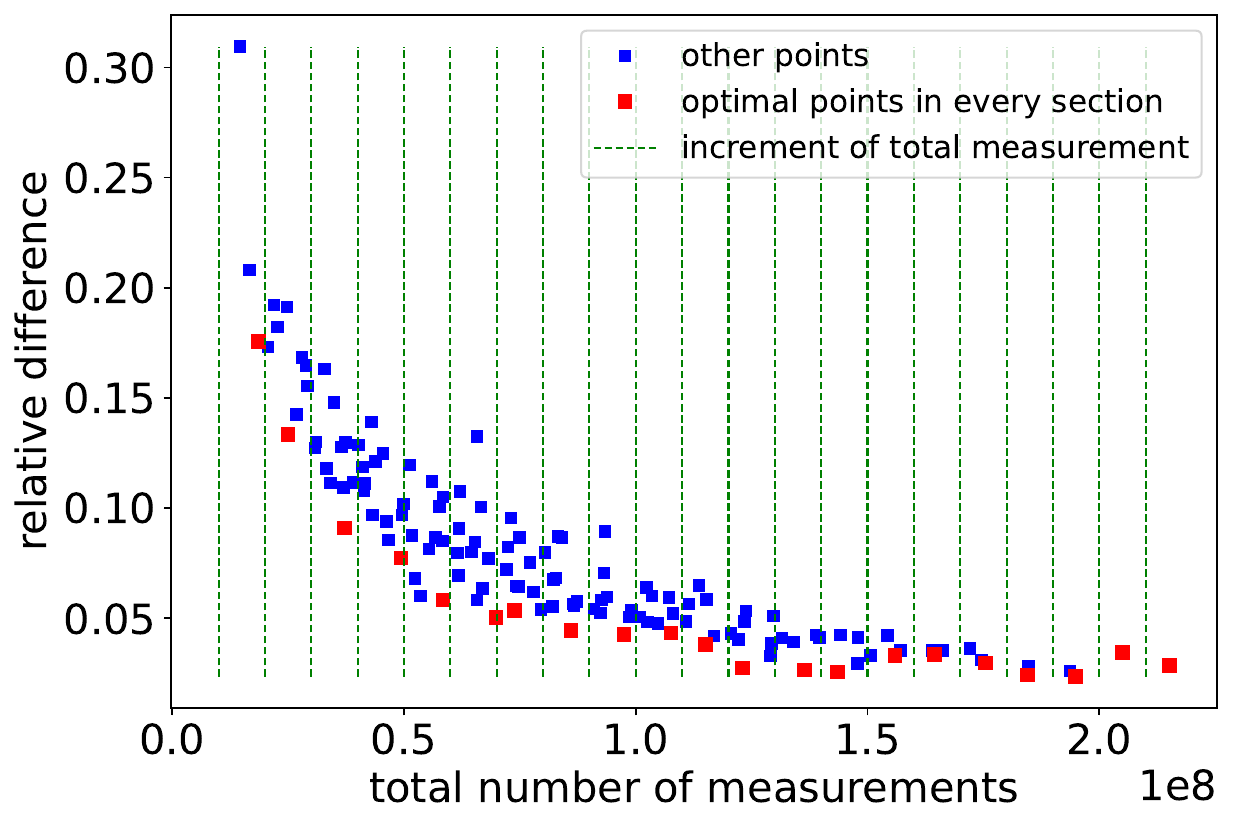}
        \label{fig:optimal parameter wins 8qd4}
    }
    \caption{Optimal choices of measurement hyperparameters ($n_{\textrm{shots}}$, $n_{\textrm{obs}}$) for (a-b) a 4-qubit depth-2 ansatz and (c-d) an 8-qubit depth-4 ansatz. The left panels show the optimal choice of hyperparameters for each interval of total number of measurements of each system size. The right panels show their final optimized energy error averaged over the final energies of 3 independent UBOS runs with different random initial states. The optimal combinations of $n_{\textrm{obs}}$ and $n_{\textrm{shots}}$ helps the algorithm reach a few percent closer to the optimal VQE energy than the non-optimal choices.}
    \label{figs:optimal choice of hyperparameters}
\end{figure*}

We consider the optimal choice of measurement hyperparameters for Edg-UBOS, $n_{\textrm{shots}}^{Edg}$ and $n_{\textrm{obs}}$, given roughly the same total amount of measurement (see Fig.~\ref{figs:optimal choice of hyperparameters}). we find that, in the range studied, $n_{\textrm{shots}}^{Edg}$ and $n_{\textrm{obs}}$ have no priority over each other so it's better to increase both hyperparameters following an alternating pattern to minimize the relative energy error to the optimal VQE energy. The difference in relative energy error to the best VQE energy between the optimal choice and the non-optimal choice is less than 5\%, which implies some flexibility in hyperparameter tuning.

\end{document}